\documentclass[11pt]{article}
\usepackage{amsfonts}
\usepackage{amsmath,amssymb,amscd}
\usepackage{stmaryrd}
\usepackage[dvips]{color}
\usepackage{fancybox}
\usepackage{bm}
\usepackage[all]{xy}
\usepackage{wrapfig}

\baselineskip=\normalbaselineskip

\def\rc#1{\textcolor{red}{#1}}

\def\Today{\ifcase\month\or January\or February\or March\or April\or May\or
 June\or July\or August\or September\or October\or November\or
December\fi\space\number\day, \number\year(\number\time)}

\def\TM{\widetilde{M}}

\def\condition#1
{\medskip\noindent{\bf Condition\ } #1 :\ }

\def\and{\quad{\rm and}\quad}

\def\p{\partial}

\def\mat#1{\left(\matrix{#1}\right)}

\def\IR{\relax{\rm I\kern-.18em R}}

\def\pr0#1#2#3{{\it Phys.\ Rev.} {\bf #1} (#2) #3}

\def\complex{{\mathchoice
{\setbox0=\hbox{$\displaystyle\rm C$}\hbox{\hbox to0pt
{\kern0.4\wd0\vrule height0.9\ht0\hss}\box0}}
{\setbox0=\hbox{$\textstyle\rm C$}\hbox{\hbox to0pt
{\kern0.4\wd0\vrule height0.9\ht0\hss}\box0}}
{\setbox0=\hbox{$\scriptstyle\rm C$}\hbox{\hbox to0pt
{\kern0.4\wd0\vrule height0.9\ht0\hss}\box0}}
{\setbox0=\hbox{$\scriptscriptstyle\rm C$}\hbox{\hbox to0pt
{\kern0.4\wd0\vrule height0.9\ht0\hss}\box0}}}}
\def\Co{{\mathchoice
{\setbox0=\hbox{$\displaystyle\rm C$}\hbox{\hbox to0pt
{\kern0.4\wd0\vrule height0.9\ht0\hss}\box0}}
{\setbox0=\hbox{$\textstyle\rm C$}\hbox{\hbox to0pt
{\kern0.4\wd0\vrule height0.9\ht0\hss}\box0}}
{\setbox0=\hbox{$\scriptstyle\rm C$}\hbox{\hbox to0pt
{\kern0.4\wd0\vrule height0.9\ht0\hss}\box0}}
{\setbox0=\hbox{$\scriptscriptstyle\rm C$}\hbox{\hbox to0pt
{\kern0.4\wd0\vrule height0.9\ht0\hss}\box0}}}}
\def\Rl{{\mathchoice
{\setbox0=\hbox{$\displaystyle\rm R$}\hbox{\hbox to0pt
{\kern0.4\wd0\vrule height0.9\ht0\hss}\box0}}
{\setbox0=\hbox{$\textstyle\rm R$}\hbox{\hbox to0pt
{\kern0.4\wd0\vrule height0.9\ht0\hss}\box0}}
{\setbox0=\hbox{$\scriptstyle\rm R$}\hbox{\hbox to0pt
{\kern0.4\wd0\vrule height0.9\ht0\hss}\box0}}
{\setbox0=\hbox{$\scriptscriptstyle\rm R$}\hbox{\hbox to0pt
{\kern0.4\wd0\vrule height0.9\ht0\hss}\box0}}}}

\begin{document}

\numberwithin{equation}{section}
\renewcommand{\theequation}{\thesection.\arabic{equation}}
\def\natural{\mathbb{N}}
\def\mat#1{\matt[#1]}
\def\matt[#1,#2,#3,#4]{\left(%
\begin{array}{cc} #1 & #2 \\ #3 & #4 \end{array} \right)}
\def\hq{\hat{q}}
\def\hp{\hat{p}}
\def\hx{\hat{x}}
\def\hk{\hat{k}}
\def\hw{\hat{w}}
\def\hl{\hat{l}}

\def\bea#1\ena{\begin{align}#1\end{align}}
\def\nn{\nonumber\\}
\def\cL{{\cal L}}
\def\TM{TM\oplus T^*M}
\newcommand{\CouB}[2]{\left\llbracket #1,#2 \right\rrbracket}
\newcommand{\pair}[2]{\left\langle\, #1, #2\,\right\rangle}

\null \hfill Preprint TU-982  \\[3em]
\begin{center}
{\LARGE \bf{
Topological T-duality via Lie algebroids and $Q$-flux in Poisson-generalized geometry 
}}
\end{center}

\begin{center}
{T. Asakawa${}^\sharp$\footnote{e-mail: asakawa@maebashi-it.ac.jp}, 
H. Muraki${}^{\flat}$\footnote{e-mail: hmuraki@tuhep.phys.tohoku.ac.jp}, 
and S. Watamura${}^{\flat}$\footnote{e-mail: watamura@tuhep.phys.tohoku.ac.jp}}\\[3em] 
${}^\sharp$
Department of Integrated Design Engineering,\\
Faculty of Engineering,\\
Maebashi Institute of Technology\\
Maebashi, 371-0816, Japan \\[1em]

${}^\flat$
Particle Theory and Cosmology Group \\
Department of Physics \\
Graduate School of Science \\
Tohoku University \\
Aoba-ku, Sendai 980-8578, Japan \\ [5ex]

\thispagestyle{empty}

\abstract{
It is known that the topological T-duality exchanges $H$ and $F$-fluxes. 
In this paper, we reformulate the topological T-duality as an exchange of two Lie algebroids 
in the generalized tangent bundle.
Then, we apply the same formulation to the Poisson-generalized geometry, 
which is introduced in \cite{AMSW} 
to define $R$-fluxes as field strength associated with $\beta$-transformations.
We propose a definition of $Q$-flux associated with $\beta$-diffeomorphisms, 
and show that the topological T-duality exchanges $R$ and $Q$-fluxes.
} 
\end{center}

\vskip 2cm

\eject

\section{Introduction}

In the effective theory of string, there are various 
types of fluxes. 
Some of the fluxes are known to appear in the analysis on the string spectrum of the worldsheet theory.
In particular, $H$-fluxes and geometric $F$-fluxes 
fit well in the framework of the generalized geometry \cite{Gualtieri,Hitchin,Hitchin2}.
Whereas some others of them, referred to as ``non-geometric fluxes," are expected to
appear after performing duality transformations
\cite{Kaloper,Hull:2004in,Shelton:2006fd,Dabholkar:2005ve,Grana,Halmagyi:2008dr},
but the geometrical characterizations of them are still missing.

Our main motivation for introducing a new geometric structure 
in the previous paper \cite{AMSW}
was to reveal the geometrical origin of such non-geometric fluxes. 
We proposed a variant of the generalized geometry as a candidate of 
a geometrical structure, which describes the one of the non-geometric fluxes.

The structure is based on
a Courant algebroid, $(T M)_0\oplus (T^*M)_\theta $,
defined on a Poisson manifold $M$ equipped with a Poisson tensor $\theta$ \cite{AMSW}.
In the standard generalized geometry \cite{Gualtieri,Hitchin,Hitchin2}, 
the Courant algebroid $TM\oplus T^*M$ is 
a basic object and can be considered as an extension of the Lie algebroid $TM$.
Similarly, in the variant of the generalized geometry, 
which we call Poisson-generalized geometry, the Courant algebroid 
$(TM)_0\oplus (T^*M)_\theta $ 
can be regarded as
an extension of the Lie algebroid $(T^*M)_\theta$ of the Poisson manifold.
They are dual with each other
in the sense that the roles of the tangent and the cotangent bundles are exchanged.
Apart from some differences, 
they are indeed equipped with analogous mathematical structures \cite{AMSW}.  
Hence, various concepts known in the standard generalized geometry 
such as Dirac structures, generalized Riemannian structures
can be established also in the Poisson-generalized geometry.

One of the major differences is in their symmetries.
The symmetry of the Courant algebroid $TM\oplus T^*M$ is given by
the semidirect product of the diffeomorphism and $B$-transformation,
whereas that of the the Courant algebroid $(TM)_0\oplus (T^*M)_\theta $
is given by the semidirect product of $\beta$-diffeomorphism and $\beta$-transformation.
As a result,
in a similar way as an $H$-flux is associated with the twist of $(T M)_0\oplus (T^*M)_\theta $
by the local $B$-gauge transformations \cite{Gualtieri,SeveraWeinstein},
the twist of $(TM)_0\oplus (T^*M)_\theta $ by the local $\beta$-gauge transformations
indicates the proper definition of the so-called $R$-flux, which is a
$3$-vector $R \in \Gamma (\Lambda^3TM)$, 
as a gauge field strength of local bivector gauge potentials \cite{AMSW}.

In this paper, we investigate the Poisson-generalized geometry further.
The aim of this paper 
is to propose a proper definition of another kind of non-geometric flux, the $Q$-flux, 
in the framework of Poisson-generalized geometry.
The strategy is to require consistency with the topological T-duality, 
and to define a $Q$-flux as the Poisson analogue of a geometric $F$-flux.

The notion of topological T-duality \cite{Bouwknegt:2003vb,Bouwknegt:2003zg}
is well understood in the framework of the generalized geometry 
\cite{Gualtieri,Cavalcanti:thesis,Cavalcanti:2011wu,Bouwknegt:2010zz}.
A remarkable feature of the topological T-duality is that it exchanges $H$-flux and 
geometric $F$-flux. Since the $F$-flux is defined as
the curvature $2$-form of a principal circle bundle, 
the topological T-duality provides a relation between two
generalized geometries defined on two different circle bundles.
This formulation is not suitable for our purpose, since we do not know the Poisson analogue 
of circle bundles.

Thus, in the former part of this paper, we reformulate the topological T-duality 
in the standard generalized geometry.
We demonstrate
the $S^1$-dimensional reduction of the generalized tangent bundle $TM\oplus T^*M$
and show that two Lie algebroids appear in two different ways in this setting.
Both of them are isomorphic to
the Lie algebroid $TN\oplus {\mathbb R}$ over the reduced base space
$N$ with $M=N\times S^1$.
Using this fact,
we show that the topological T-duality can be reformulated as an exchange of these two Lie algebroids.
Then we see 
that both the $2$-form part of an $H$-flux and of a geometric $F$-flux 
are associated with the twisting of the same Lie algebroid $TN\oplus {\mathbb R}$.
Thus, the topological T-duality results in the exchange of these fluxes, when fluxes are present.

An advantage of this reformulation 
of the topological T-duality from the viewpoint of Lie algebroid
is to keep the base space $M$ unchanged under the topological T-duality.
Thus, it is easy to apply the same procedure 
to the case of the Poisson-generalized geometry.
In the latter part of this paper,
we show that we can find an analogue of the topological T-duality in the $S^1$-reduction of the new Courant algebroid $(TM)_0\oplus (T^*M)_\theta $,
where the Lie algebroid $(T^*N)_\theta \oplus {\mathbb R}$ appears in two different ways.
By considering the twisting of this Lie algebroid, 
$Q$-flux can be naturally defined.
This $Q$-flux is defined as a gauge field strength bivector associated 
with $\beta$-diffeomorphisms, 
as a counterpart of the geometric $F$-flux associated with ordinary
diffeomorphisms in the standard generalized geometry.
We then show the consistency of our $R$ and $Q$-flux with the topological T-duality,
which is summarized as the exchange of the bivector parts of $R$-flux and $Q$-flux.

The organization of this paper is as follows:
In \S 2, we recall the basic setting of both the standard generalized geometry
and the Poisson-generalized geometry.
We also give characterizations of $H$ and $R$-fluxes in terms of bundle maps $\varphi$,
which is useful to define twisted brackets and 
to give a simple proof of topological T-duality in the subsequent sections. 
In \S 3, we demonstrate
the $S^1$-dimensional reduction of the generalized tangent bundle $TM\oplus T^*M$
and reformulate the topological T-duality 
using the Lie algebroid $TN\oplus {\mathbb R}$.
By considering $H$ and $F$-fluxes as twistings of $TN\oplus {\mathbb R}$,
we recover the exchange of $H$ and $F$-fluxes in this formulation.
Then in \S 4, we apply the topological T-duality 
in the new formulation in \S 3  
to the case of the Poisson-generalized geometry $(TM)_0\oplus (T^*M)_\theta $.
After formulating 
the topological T-duality using the Lie algebroid $(T^*N)_\theta \oplus {\mathbb R}$,
we propose a geometrical definition of $Q$-fluxes.
Then, we show that $R$ and $Q$-fluxes are also exchanged 
by the topological T-duality.
\S 5 is devoted to the conclusion and discussion.

\section{Generalized geometry and Poisson-generalized geometry}

We first recall the basic setting of the generalized and the Poisson-generalized geometries.
See \cite{AMSW} in more detail.

\subsection{Generalized geometry}
The generalized geometry is formulated in terms of the Courant algebroid $TM\oplus T^*M$, the generalized tangent bundle, with the inner product,  the anchor map 
and the Courant bracket
being defined as, for the generalized tangent vectors $e_1=X+\xi$ and $e_2=Y+\eta$,
\bea
\langle  e_1,  e_2 \rangle&= \textstyle{\frac{1}{2}}(i_X \eta+i_Y \xi),
~~ \rho(e_1) =X,
\\ [e_1,e_2]_C &=[X,Y]+{\cal L}_X \eta -{\cal L}_Y \xi -\textstyle{\frac{1}{2}}d(i_X\eta-i_Y\xi),
\label{Courant bracket}
\ena
respectively.
The symmetry of the Courant algebroid consists 
of diffeomorphisms and $B$-transformations.
These transformations can be represented by using a vector $Z \in \Gamma(TM)$
 and a 2-from $b\in \Gamma( \wedge^2 T^*M)$ with $db=0$, as matrices
\bea
\mat{e^{{\cal L}_Z},0,0,e^{{\cal L}_Z}}, ~~\mat{1,0,b,1},
\ena
where each matrix element is a bundle map and the matrices are acting on 
the column vector $(\Gamma(TM),\Gamma(T^*M))^t$ in the representation space.

An $H$-flux is specified by the data $(H,B_i,A_{ij})
 \in \Gamma(\wedge^3 T^*M) \times \Gamma(\wedge^2 T^*U_i) \times \Gamma(T^*U_{ij})$
 such that
\bea
H|_{U_i}=dB_i, ~~B_j-B_i|_{U_{ij}} =dA_{ij},
\ena
where $\{U_i\}$ is a good open covering of $M$, and $U_{ij}=U_i \cap U_j$.
An $H$-twisting of $TM\oplus T^*M$ is a construction of a new Courant algebroid $E$ from the data, 
satisfying the exact sequence
\bea
0\to T^*M \xrightarrow{\rho^*} E \overset{\rho}{\underset{s}{\rightleftarrows}} TM \to 0,
\label{H-exact sequance}
\ena
with a splitting $s:TM\to E$, where $\rho$ denotes the anchor map $E \to TM$.
The splitting $s$ is given locally by the $B$-transformation of $X\in \Gamma(TM)$ as
\bea
s(X)=X+B_i (X).
\ena
The Courant algebroid $E$ is written\footnote{
We omit the symbol $\rho^*$ of the inclusion for notational simplicity.}
 as $E=s(TM)\oplus T^*M$.
In this paper, we also denote the splitting as a bundle map 
$\varphi_H=s\oplus {\rm id.} :TM\oplus T^*M \to E$.
The map $\varphi_H$ is globally-defined, and can be locally represented by a matrix
\bea
\varphi_H  =\mat{1,0,B_i,1}.
\label{phi H}
\ena
Then, the $H$-twisted bracket on $TM\oplus T^*M$ can be defined by 
\bea
[e_1,e_2]_H :=\varphi_H^{-1}[\varphi_H (e_1), \varphi_H (e_2)]_C.
\ena
Substituting $e_1=X+\xi$ and $e_2=Y+\eta$, the above $H$-twisted bracket 
 gives
\bea
[X+\xi,Y+\eta]_H =[X+\xi,Y+\eta]_C-i_{X}i_{Y}H.
\ena
It is shown that 
the Courant algebroid $E$ with the Courant bracket and $TM\oplus T^*M$ with the 
$H$-twisted bracket are isomorphic to each other.

\subsection{Poisson-generalized geometry}
The Poisson-generalized geometry, introduced in \cite{AMSW}, is formulated in terms of the Courant algebroid 
$(TM)_0\oplus (T^*M)_\theta$, with the inner product and the anchor map
\bea
\langle  e_1,  e_2 \rangle= \textstyle{\frac{1}{2}}(i_\xi Y +i_\eta X ),
~~\rho(e_1) =\theta (\xi),
\ena
for generalized tangent vectors $e_1=X+\xi$ and $e_2=Y+\eta$, and the bracket 
\bea
[e_1,e_2] =[\xi,\eta]_\theta +{\cal L}_\xi Y -{\cal L}_\eta X 
-\textstyle{\frac{1}{2}}d_\theta (i_\xi Y -i_\eta X).
\label{Sasa bracket}
\ena
Here, ${\cal L}_\xi$, $d_\theta$ and $i_\xi$ 
are the $A$-Lie derivative, the $A$-differential and the $A$-interior product 
of the Lie algebroid $A=(T^*M)_\theta$ of the Poisson manifold, respectively.
The bracket $[\cdot,\cdot]_\theta$ is the Lie bracket of the Lie algebroid 
$(T^*M)_\theta$, which is so-called the Koszul bracket.
The symmetry of this Courant algebroid consists of $\beta$-diffeomorphisms and $\beta$-transformations.
For a $1$-from $\zeta \in \Gamma(T^*M)$ with ${\cal L}_\zeta \theta =0$,
and a bi-vector $\beta \in \Gamma (\wedge^2 T M)$ with $
d_\theta \beta=0$, they are represented as
\bea
\mat{e^{{\cal L}_\zeta},0,0,e^{{\cal L}_\zeta}}, ~~\mat{1,\beta,0,1},
\ena
as a bundle map acting on $(\Gamma(TM),\Gamma(T^*M))^t$.

An $R$-flux is specified by the data 
$(R,\beta_i,\alpha_{ij}) \in \Gamma(\wedge^3 TM) \times \Gamma(\wedge^2 TU_i)
 \times \Gamma(TU_{ij})$ such that
\bea
R|_{U_i}=d_\theta \beta_i, ~~\beta_j-\beta_i|_{U_{ij}} =d_\theta \alpha_{ij}.
\ena
where $\{U_i\}$ is a good open covering of $M$, and $U_{ij}=U_i \cap U_j$.
An $R$-twisting of $(TM)_0\oplus (T^*M)_\theta$ is a construction 
of a new Courant algebroid $E$ from the data, satisfying the exact sequence
\bea
0\to (TM)_0 \xrightarrow{\pi^*} {E}
 \overset{\pi}{\underset{s}{\rightleftarrows}} (T^*M)_\theta \to 0,
\label{R-exact sequance}
\ena
with a splitting $s:(T^*M)_\theta \to {E}$,
where $\pi:E \to (T^*M)_\theta$ denotes the canonical projection (not the anchor map).
Here the splitting $s$ is given locally 
\bea
\sigma(\xi )=\xi +\beta_i (\xi),
\ena
and the Courant algebroid ${E}$ is written as 
${E}=(T M)_0 \oplus s((T^*M)_\theta ) $.
We denote this splitting as a bundle map 
$\varphi_R={\rm id.}\oplus s :(TM)_0\oplus (T^*M)_\theta \to {E}$, 
which is locally given by 
\bea
\varphi_R  =\mat{1,\beta_i,0,1}.
\label{phi R}
\ena
By using this, the $R$-twisted bracket on $(TM)_0\oplus (T^*M)_\theta$ is defined by 
\bea
[e_1,e_2]_R :={\varphi_R^{-1}}[\varphi_R (e_1), \varphi_R (e_2)],
\ena
and it is calculated for $e_1=X+\xi$ and $e_2=Y+\eta$ as
\bea
[e_1,e_2]_R =[e_1,e_2] -i_{\xi}i_{\eta}R.
\ena
It is shown that the Courant algebroid ${E}$ with the bracket (\ref{Sasa bracket}) 
and $(TM)_0\oplus (T^*M)_\theta$ with the 
$R$-twisted bracket are isomorphic to each other.

\section{Topological T-duality of Generalized Geometry}

In this section, we discuss a property of the topological T-duality 
in the standard generalized geometry, by using 
$S^1$-dimensional reduction of a target manifold $M=N\times S^1$.
Although the concepts of the topological T-duality are well-known and explained already in the framework 
of generalized geometry in references 
\cite{Gualtieri}\cite{Cavalcanti:thesis}\cite{Cavalcanti:2011wu}\cite{Bouwknegt:2010zz},
we here present a new formulation of the topological T-duality in terms of the Lie algebroid.
In this formulation, the following 
aspects of the topological T-duality become manifest:
\begin{enumerate}
\item[1)] A Lie algebroid $TN\oplus {\mathbb R}$ appears in 
the generalized tangent bundle $TM\oplus T^*M$ in two different ways, namely as
$TN \oplus \langle \p_y \rangle $ and $TN \oplus \langle dy \rangle $.
The topological T-duality in the absence of fluxes
is formulated as an exchange of these two Lie algebroids.
\item[2)] An $F$-flux is introduced as a field strength
associated with a twisting of the Lie algebroid 
$TN\oplus {\mathbb R}$.
\item[3)] The twist of 2) is included in $TM\oplus T^*M$ as
two different twistings, $H_2$- and $F$-twisting.
The former twist is caused by the $2$-form part of the $H$-flux, 
while the latter  by the geometric $F$-flux.
\item[4)] The topological T-duality exchanges $H_2$- and $F$-fluxes.
\end{enumerate}
The notations above and the details will be explained in the following subsections.

\subsection{Topological T-duality without flux}

Let us assume that a target manifold $M$ is 
a direct product\footnote{
Our presentation is valid also for $M=N\times {\mathbb R}$.} $M=N\times S^1$, 
with local coordinates $(x^m,y)$, and regard $M$ 
as a trivial $S^1$-bundle over the base space $N$.
Then, the $S^1$-dimensional reduction of the tangent bundle $TM$
corresponds to restricting the space of vector fields $\Gamma(TM)$ 
to the $S^1$-invariant vector fields, which we call {\it basic} throughout this paper.
The basic vector field has  the form 
\bea
X=X_1+f\p_y,
\ena
where $X_1 =X^m (x) \p_m \in \Gamma(TN)$ and $f (x) \in C^\infty (N)$ are a vector field and 
a function on $N$, respectively.
Note that they are independent of local $S^1$-coordinate $y$.
We denote the space of basic vector fields as $\Gamma (TM)_{\rm basic}$.
The tangent bundle $TM$ is a Lie algebroid $(TM,\rho={\rm id.},[\cdot,\cdot])$,
with the anchor map $\rho$ being the identity map, and with the Lie bracket of vector fields.
The space of the basic vector fields $\Gamma (TM)_{\rm basic}$ closes under the Lie bracket
\bea
[X,Y]&= [X_1+f\p_y, Y_1+g\p_y]
=[X_1,Y_1]_{TN} + \left({\cal L}_{X_1} g -{\cal L}_{Y_1}f \right)\p_y~.
\label{KK-reduced TM}
\ena

This bracket is the same as the Lie bracket of the Lie algebroid 
$A=TN\oplus {\mathbb R}$ over $N$, 
where ${\mathbb R}$ denotes the trivial line bundle over $N$.
In the Lie algebroid $A=TN\oplus {\mathbb R}$, 
the elements have the form $X_1 +f \in TN\oplus {\mathbb R}$, the anchor map 
$\rho_A: A \to TN$ is defined by
$\rho_A (X_1+f)=X_1$, 
and its Lie bracket is given by
\bea
[X_1+f, Y_1+g]_A
=[X_1,Y_1]_{TN} + \left({\cal L}_{X_1} g -{\cal L}_{Y_1}f \right).
\label{bracket of TN+R}
\ena
Thus the dimensional reduction reduces the Lie algebroid $TM$ over $M$ to 
$TN\oplus {\mathbb R}$ over $N$,
with the identification of the anchor map $\rho_A=\rho|_{TN}$, 
the restriction of the anchor map of $TM$ to $TN$.
To distinguish with another Lie algebroid given below, we denote 
the above Lie algebroid as 
$TN \oplus \langle \p_y\rangle$, 
and thus $\Gamma (TM)_{\rm basic}\simeq \Gamma(TN \oplus \langle \p_y\rangle )$.

The same Lie algebroid $A=TN\oplus {\mathbb R}$ 
can appear differently in the generalized geometry $TM\oplus T^*M$.
Consider a subbundle $L={\rm span}\{\p_m, dy\} $ over $M$, 
which is a Dirac structure in the generalized tangent bundle $TM\oplus T^*M$. 
In general, a Dirac structure is a Lie algebroid with respect to the Courant bracket (\ref{Courant bracket}).
The $S^1$-dimensional reduction
restricts the space of the sections $\Gamma(L)$ 
to the basic sections $\Gamma (L)_{\rm basic}$ of the form 
\bea
X_1+f dy,
\ena
where $X_1 \in \Gamma(TN)$ and $f (x) \in C^\infty (N)$ are basic, i.e., independent of 
 the local $S^1$-coordinate $y$. 
The Courant bracket of the basic sections of $L$ gives
\bea
[X_1+fdy, Y_1+gdy]_C
=[X_1,Y_1]_{TN} + \left({\cal L}_{X_1} g -{\cal L}_{Y_1}f \right) dy,\label{CourantBasicdy}
\ena
which is identical to (\ref{bracket of TN+R}), the Lie bracket of $A=TN\oplus {\mathbb R}$.
We denote this Lie algebroid as $TN \oplus \langle dy\rangle$, 
and thus $\Gamma (L)_{\rm basic}\simeq \Gamma(TN\oplus \langle dy\rangle ) $.

In summary, the Lie algebroid $A=TN\oplus {\mathbb R}$ appears in two different ways 
in the framework of $TM\oplus T^*M$, and the bundle map 
$\mathcal{T}: TN\oplus \langle \p_y\rangle \to TN\oplus \langle dy\rangle$
defines a Lie algebroid  isomorphism, which induces 
the map between the sections as
\bea
\mathcal{T}: X_1+f\p_y \mapsto X_1+fdy.
\ena
The situation can be summarized schematically as follows:  
\begin{align}
\xymatrix{
& TM\oplus T^*M \ar[ld]\ar[rd] & \\
TN\oplus \langle \p_y\rangle  \ar[rr]^\simeq && TN\oplus \langle dy\rangle ,
}
\end{align}
where each diagonal arrow represents the 
dimensional reduction to the corresponding bundle over $N$.
Note that the left diagonal arrow is 
accompanied by the restriction of the anchor map. The horizontal arrow represents the map 
$\mathcal{T}$.

This isomorphism is the key of the topological T-duality.
In fact, the map $\mathcal{T}$ can be extended to the automorphism of the $S^1$-reduced 
generalized tangent bundle $TN\oplus \langle \p_y\rangle \oplus T^*N\oplus \langle dy\rangle$.
For the basic sections $\Gamma (TM\oplus T^*M)_{\rm basic}$ of the form
\bea
e=X_1+f\p_y +\xi_1 +h dy,
\ena
where $X_1 \in \Gamma(TN)$, $\xi_1 \in \Gamma(T^*N)$ and $f , h\in C^\infty (N)$,
the extension of the map $\mathcal{T}$ is given by
\bea
\mathcal{T}:X_1+ f\p_y +\xi_1 +h dy \mapsto X_1+ h\p_y +\xi_1 +f dy~,
\ena
which just yields the interchange of $f$ and $h$.
Then, it can be shown (see appendix \ref{autoofT}) that the map $\mathcal{T}$ preserves 
the inner product, anchor map (restricted to $TN$), and the
Courant bracket
\bea
&\langle  \mathcal{T}e_1,  \mathcal{T}e_2 \rangle= \langle  e_1, e_2\rangle ,
 ~~\rho|_{TN} ( \mathcal{T}e) =\rho|_{TN} (e), \nn
&[\mathcal{T}e_1, \mathcal{T}e_2]_C= \mathcal{T}[e_1, e_2]_C~. \label{T preserves Coutant}
\ena
Thus, the map $\mathcal{T}$ defines an automorphism.
In other words, $\mathcal{T}$ defines an extra symmetry valid only for the basic sections.

\subsection{$F$-twisting of the Lie algebroid $TN\oplus {\mathbb R}$}

It is known\footnote{The content of this section is a well-known material as Atiyah algebroids. 
See for example \cite{Gualtieri,Rogers,CdSWeinstein}.}
that the twisting of $TN\oplus {\mathbb R}$ gives another Lie algebroid $A$ over $N$,
satisfying the exact sequence
\bea
0\to {\mathbb R} \to A \to TN \to 0,
\ena
and which is classified by $[F] \in H_{\rm dR}^2(N)$ in the de Rham cohomology.
The corresponding bracket in $TN\oplus {\mathbb R}$ becomes an $F$-twisted one:
\bea
[X_1+f, Y_1+g]_F
=[X_1,Y_1]_{TN} + \left({\cal L}_{X_1} g -{\cal L}_{Y_1}f \right) -F(X_1,Y_1).
\label{F-twisted bracket}
\ena

The procedure to obtain the $F$-twisted bracket is analogous to the case of the $H$-twist.
Given a good cover $\{U_i\}$ of $N$ and a trivialization $(F,a_i,\lambda_{ij})$ of a closed $2$-form $F$
such that
\bea
F~\big|_{U_i}=da_i, \quad 
a_j -a_i ~\big|_{U_{ij}}=d\lambda_{ij},
\ena 
we define a transition function $G_{ij}:U_{ij}\to GL(d)$ ($d=\dim N+1 =\dim M$) by
\bea
G_{ij}=\mat{1,0,d\lambda_{ij},1}.
\label{bareF transition}
\ena
Then, since $G_{ij}$ satisfies the cocycle condition, we obtain a vector bundle  over $N$
\bea
A=\coprod_{i\in N} TU_i \oplus {\mathbb R}_i /\sim.
\ena
Moreover, 
$A$ is a 
Lie algebroid, since each $TU_i \oplus {\mathbb R}_i \to U_i$ is a Lie algebroid and 
the gluing condition preserves the anchor map and the Lie bracket.
The set $\{a_i\}$ of local $1$-forms gives a splitting $\tilde{s}:TN\to A$, locally defined by
\bea
\tilde{s}(X_1)=X_1-a_i (X_1),
\ena
for $X_1 \in TU_i$.
Since $\tilde{s}(TN)$ is globally well-defined, any section of $A$ is uniquely specified by
\bea
\tilde{s}(X_1)+f ,
\ena
for $X_1 \in \Gamma (TN)$ and $f \in C^\infty (N)$.
The Lie bracket of these sections can be calculated straightforwardly as
\bea
[\tilde{s}(X_1)+f , \tilde{s}(Y_1)+g ]_A 
=\tilde{s} ([X_1, Y_1])+{\cal L}_{X_1}g-{\cal L}_{Y_1}f -F(X_1,Y_1),	\label{Liebracket318}
\ena
which is identified with the $F$-twisted Lie bracket (\ref{F-twisted bracket}) 
under the identification of $A= \tilde{s}(TN)\oplus {\mathbb R}$ 
and $TN\oplus {\mathbb R}$.

It is obvious from the construction that the $2$-form flux $F$ is a field strength 
associated with an abelian gauge symmetry defined by a set of local gauge parameters 
$f_i \in C^\infty (U_i)$ where the gauge transformation of the data $(F,a,\lambda)$
is given by\footnote{It is valid for both gauge group $G={\mathbb R}$ and $G=U(1)$.
For $G=U(1)$, 
it can be written as more familiar form $g_{ij}\mapsto e^{i f_j}g_{ij}e^{-if_i}$ for $g_{ij}=e^{i\lambda_{ij}}$.}
\bea
F \mapsto F,~~ a_i \mapsto a_i +df_i, ~~\lambda_{ij} \mapsto \lambda_{ij} +f_j -f_i~.
\ena
If $[F/2\pi]$ is the image of the map of cohomologies
$H^2(N;{\mathbb Z}) \to H^2(N;{\mathbb R})\simeq H_{\rm dR}^2(N)$, i.e., the 1st Chern class, 
then $A$ is identified with the Atiyah algebroid $TP/U(1)$, 
where $P$ is a principal $S^1$-bundle over $N$
with connection.

\subsection{$H$ and $F$-fluxes in generalized geometry}

The $F$-twisting of Lie algebroid $TN\oplus {\mathbb R}$ above
can also appear in two different twistings of the dimensional reduction of the 
generalized tangent bundle $TM\oplus T^*M$:
\begin{enumerate}
\item
By the $F$-twisting of $TN\oplus \langle \p_y\rangle$, 
the $F$-twisted bracket (\ref{F-twisted bracket}) corresponds to
\bea
[X_1+f\p_y, Y_1+g\p_y]_F
=[X_1,Y_1]+ \left( \left({\cal L}_{X_1} g -{\cal L}_{Y_1}f \right) -F(X_1,Y_1)\right) \p_y.
\label{geometric F-twisted bracket}
\ena
In this case $F$ is called a geometric flux, since this Lie algebroid is equivalent to 
a principal $S^1$-bundle $P$ over $N$ with curvature $F$.
\item
By the $F$-twisting of $TN\oplus \langle dy\rangle$,
the $F$-twisted bracket (\ref{F-twisted bracket}) corresponds to
\bea
[X_1+fdy, Y_1+gdy]_F
=[X_1,Y_1]+ \left( \left({\cal L}_{X_1} g -{\cal L}_{Y_1}f \right) -F(X_1,Y_1)\right) dy.
\ena
This twisting is necessarily a part of an $H$-twisting, 
since the twisting of $\p_m$ and $dy$ is achieved by a $B$-transformation.
We denote such $2$-form $F$ as $H_2 \in \wedge^2 T^*N$.
Then, the corresponding $3$-form $H$-flux is $H=-H_2 \wedge dy$.
\end{enumerate}
We may consider more generally twistings of basic sections of the generalized tangent bundle 
by both an $H$-flux and an $F$-flux, represented schematically as
\begin{align}
\xymatrix{
& TM\oplus T^*M  \ar[ld]\ar[d]\ar[rd] & \\
TN\oplus \langle \p_y\rangle \ar@(ld,rd)_{F} 
&TN\oplus T^*N \ar@(ld,rd)_{H_3}
& TN\oplus \langle dy\rangle \ar@(ld,rd)_{H_2},
}
\end{align}
Here $2$-forms $F$ and $H_2$ on $N$ correspond to the $F$-twistings described above,
and the $3$-form $H_3$ on $N$ corresponds to the $H$-twisting of $TN\oplus T^*N$.

In the following, we formulate this schematic picture  
more precisely in the framework of the Courant algebroid.
To this end, first we twist $TN\oplus \langle \p_y \rangle$ by $F$ to obtain a 
Courant algebroid $A\oplus A^*$ 
over $N$ with $A=\tilde{s}(TN) \oplus \langle \p_y\rangle$.
Then, we twist $A\oplus A^*$ by $H$ to obtain a Courant algebroid 
$E=s(A)\oplus A^*$, which is equivalent to the 
$S^1$-reduced generalized tangent bundle with $(H,F)$-twisted Courant bracket.

\subsubsection{Geometric $F$-flux}

By the $F$-twisting of the Lie algebroid $TN\oplus \langle \p_y\rangle$ 
for given data $(F,a_i,\lambda_{ij})$, 
we obtain a Lie algebroid $A\simeq \tilde{s}(TN)\oplus \langle \p_y \rangle$,
whose section has the form
\bea
X=\tilde{s}(X_1)+f\p_y, \quad \tilde{s}(X_1)=X_1-a_i (X_1)\p_y.		\label{323}
\ena
If we regard $A=TP/U(1)$, 
where $P$ is a principal $S^1$-bundle over $N$ with curvature $F$,
$\tilde{s}(X_1)$ defines the horizontal lift of $X_1$ in $TP$, while $\p_y$ is the vertical direction. 
Considering its dual $A^*=T^*P/U(1)$, 
the corresponding connection $1$-form $a$ on $P$ is 
defined locally by
\bea
a|_{U_i}=dy+a_i ,
\label{def of A}
\ena
and an arbitrary basic $1$-form $\xi$ on $M$ is decomposed with respect to $a$ as
\bea
\xi=\xi_1 +h a,
\ena
where $\xi_1 \in \Gamma(T^*N)$ and $h \in C^\infty (N)$.
As usual, the following relations hold:
\bea
i_{\tilde{s}(X_1)}a=0, \quad i_{\p_y}a=1, 
\quad i_{\tilde{s}(X_1)}\xi_1=i_{X_1}\xi_1, \quad i_{\p_y}\xi_1=0.	\label{326}
\ena
With this in mind, any section of the $F$-twisted Courant algebroid 
$A\oplus A^*$ is written of the form
\bea
\tilde{s}(X_1)+f\p_y +\xi_1 +ha.
\ena
Therefore, corresponding to the representation of the twisting in \S 2, 
the $F$-twisting is specified by a bundle map
$\varphi_F:TN\oplus \langle \p_y\rangle \oplus 
T^*N \oplus \langle dy\rangle \to A\oplus A^*$,
\bea
\varphi_F  =\mat{e^{-a_i \wedge \p_y},0,0,e^{-a_i \wedge \p_y}},
\label{F map}
\ena
locally defined on $U_i$.
Indeed, it is easily seen that
\bea
\varphi_F (X_1 +f\p_y+\xi_1 +hdy)=\tilde{s}(X_1)+f\p_y +\xi_1 +ha,
\ena
because
\bea
&e^{-a_i \wedge \p_y}(X_1 +f\p_y )=X_1 +f\p_y -a_i (X_1) \p_y =\tilde{s}(X_1)+f\p_y,\nn
&e^{-a_i \wedge \p_y}(\xi_1 +hdy)=\xi_1 +hdy +h a_i =\xi_1 +ha.
\ena
We emphasize that the $F$-twisting in the previous subsection affects both on $TN\oplus \langle \p_y\rangle$ and 
$T^*N \oplus \langle dy\rangle$, diagonally.

Corresponding to (\ref{F map}), the transition function (\ref{bareF transition}) is also embedded diagonally as
\bea
G^F_{ij} =\mat{e^{-d\lambda_{ij} \wedge \p_y},0,0,e^{-d\lambda_{ij} \wedge \p_y}}.
\ena 
It is a diffeomorphism $e^{{\cal L}_{Z_{ij}}}$ 
generated by the vector field $Z_{ij}=\lambda_{ij} \p_y \in \langle \p_y \rangle$ 
on $U_{ij}$.\\
{\it Proof.} We see that the action of ${\cal L}_{Z_{ij}}$ is equivalent to the gluing condition.
\bea
&{\cal L}_{Z_{ij}}(X_1 +f\p_y )=-({\cal L}_{X_1}\lambda_{ij} )\p_y =-(i_{X_1}d\lambda_{ij})\p_y
=-d\lambda_{ij}(X_1) \p_y ,\nn
&{\cal L}_{Z_{ij}}(\xi_1 +hdy)=h ({\cal L}_{Z_{ij}}dy) =h(di_{Z_{ij}} dy) =hd\lambda_{ij}.
\ena
{\it (End of the proof.)}\\
This shows that $F$-fluxes are associated with diffeomorphisms, and thus called geometric fluxes\footnote
{When considering a generalized metric, this diffeomorphism is seen as the Kaluza-Klein $U(1)$-gauge symmetry.}.

The $F$-twisted bracket on $TN\oplus \langle \p_y\rangle \oplus T^*N \oplus \langle dy\rangle$ is defined by 
using (\ref{F map}) as
\bea
[e_1,e_2]_F :=\varphi_F^{-1}[\varphi_F (e_1), \varphi_F (e_2)]_C.
\ena
By direct calculation, we see that
the vector part of this bracket is indeed (\ref{geometric F-twisted bracket}).

\subsubsection{($H$,$F$)-flux}
By twisting $A\oplus A^*$ 
with $H_2$ and $H_3$ further, we obtain a Courant algebroid $E$, 
specified by a bundle map $\varphi_{H} : A \oplus A^* \to E$.
Hence, the twisting of $TN\oplus \langle \p_y\rangle \oplus T^*N \oplus \langle dy\rangle$ 
by both $H$ and $F$-flux is specified by the bundle map
\bea
\varphi_{H,F} := \varphi_H \varphi_F
&=\mat{1,0,B_i,1}\mat{e^{-a_i \wedge \p_y},0,0,e^{-a_i \wedge \p_y}}.
\label{phi HF}
\ena
Here the set of local $2$-forms $\{B_i\}$ defines a splitting
$s: A \to E$, and $A$ already carries the information of the $F$-flux.
Correspondingly, any basic $k$-form on $M$, say $\omega$,
is decomposed with respect to $a$ in (\ref{def of A}) as 
\bea
\omega  =\omega_k +\omega_{k-1}\wedge a,
\ena
where $k$ and $k-1$ are degrees as forms on $N$.
In particular, we decompose the $H$-flux into
\bea
&H=H_3-H_2\wedge a.
\label{H decomposition}
\ena
The decomposition is also applied to local differential forms defined on each open set $U_i$ of $N$.
In particular, we decompose $B_i$ in (\ref{phi HF}) into
\bea
&B_i=B_{2i} -B_{1i} \wedge a, \quad B_{2i} \in \Gamma (\wedge^2 T^*U_i ) ,~ B_{1i} \in 
\Gamma (T^*U_i).
\ena
Then, by noting 
\bea
H|_{U_i}&=dB_i =(dB_{2i} +B_{1i} \wedge F)-dB_{1i} \wedge a,
\ena
we should identify 
\bea
H_3 |_{U_i}=dB_{2i} +B_{1i} \wedge F ,\quad
H_2 |_{U_i}=dB_{1i}.
\ena
Thus, local $1$-form $B_{1i}$ is the gauge potential for the $H_2$-flux.

With this decomposition, (\ref{phi HF}) means that the section of $E$
is locally given by
\bea
\varphi_{H,F}(e)
&=\varphi_{H}(\tilde{s}(X_1)+f\p_y + \xi_1 +ha) \nn
&=\tilde{s}(X_1)+f\p_y + \xi_1 +ha + i_{\tilde{s}(X_1)+f\p_y} B_i \nn
&=\tilde{s}(X_1)+f\p_y + \xi_1 +ha + i_{\tilde{s}(X_1)+f\p_y} (B_{2i} -B_{1i} \wedge a) \nn
&=\tilde{s}(X_1)+f\p_y +(\xi_1 +i_{X_1}B_{2i} +fB_{1i} )+(h- B_{1i} (X_1)) a ,
\label{HF-twisted section}
\ena
for $e=X_1+f\p_y +\xi_1 +hdy$.

The $(H,F)$-twisted bracket on 
$TN\oplus \langle \p_y\rangle \oplus T^*N \oplus \langle dy\rangle$ is defined by 
\bea
[e_1,e_2]_{H,F} :=\varphi_{H,F}^{-1}[\varphi_{H,F}(e_1), \varphi_{H,F} (e_2)]_C.
\label{HF-twisted bracket}
\ena
By a direct calculation using (\ref{HF-twisted section}), we have explicitly
\bea
&[X_1+f\p_y +\xi_1 +hdy,  Y_1+g\p_y +\eta_1 +kdy]_{H,F} \nn
=&[X_1+f\p_y +\xi_1 +hdy,  Y_1+g\p_y +\eta_1 +kdy]_C \nn
&-i_{X_1}i_{Y_1}H_3
+(i_{X_1}i_{Y_1}F )\p_y +\left(i_{X_1}i_{Y_1}H_2\right)a
+ki_{X_1}F -hi_{Y_1}F -fi_{Y_1}H_2 +gi_{X_1}H_2.
\ena

\subsection{Topological T-duality with $H$ and $F$-fluxes}

We will see that the topological T-duality yields following relations:
\begin{align}
\xymatrix{
& TM\oplus T^*M \ar[ld]\ar[d]\ar[rd] & \\
TN\oplus \langle \p_y \rangle \ar@(ld,rd)_{\hat F} 
&TN\oplus T^*N \ar@(ld,rd)_{\hat H_3}
& TN\oplus \langle dy\rangle \ar@(ld,rd)_{\hat H_2},
}
\end{align}
Here the T-dual fluxes $(\hat H, \hat F)$ define the T-dual Courant algebroid $\hat{E}$, 
 or equivalently the $S^1$-reduced generalized tangent bundle 
$TN\oplus \langle \p_y\rangle \oplus T^*N \oplus \langle dy\rangle$ 
with the $(\hat H, \hat F)$-twisted bracket.
This twisting is governed by a map $\varphi_{\hat H,\hat F}$, the analogue of (\ref{phi HF}).
$3$-form $\hat{H}$ on $M$ is decomposed like (\ref{H decomposition}) into
\bea
\hat{H}=\hat{H}_3-\hat{H}_2 \wedge \hat{a},
\ena
where $\hat{a}$ with $\hat{F}=d\hat{a}$ is the connection 
$1$-form on the T-dual $S^1$-principal bundle $\hat P$ over $N$.

Since the T-duality interchanges $\p_y$ and $dy$, 
the T-dual fluxes $(\hat{H},\hat{F})$ are related to the original fluxes by
\bea
(\hat{H}_3,\hat{H}_2,\hat{F}) =(H_3,F,H_2).
\ena
From this we can observe the standard result, the T-duality exchanges $H_2$ and $F$ 
\cite{Bouwknegt:2003vb}\cite{Bouwknegt:2003zg}.

Note that in this presentation, the total space $M=N\times S^1$ as a manifold is unchanged by the T-duality.
In another view, we may also consider the space $M$ with an $F$-flux as a principal $S^1$-bundle $P$.
In this case the topological T-duality relates different principal $S^1$-bundles with $3$-form fluxes
$\mathcal{T}: (P,H) \to (\hat{P},\hat{H})$ as \cite{Bouwknegt:2003vb}.
Now we elaborate on the derivation of this result.

\subsubsection{Details on topological T-duality with ($H$,$F$)-flux}
The topological T-duality is defined still as a bundle map of 
$S^1$-reduced generalized tangent bundle
$TN\oplus \langle \p_y\rangle \oplus T^*N \oplus \langle dy\rangle$.
It acts on basic sections as
\bea
\mathcal{T}:X_1 +f \p_y +\xi_1 +h dy \mapsto X_1 +h\p_y +\xi_1 +f dy.
\label{T-duality rule}
\ena
After twisting, it is also regarded as a map $\mathcal{T}:E\to \hat{E}$ 
of twisted Courant algebroids.
The key relation between sections of $E$ and $\hat{E}$ is
\bea
\mathcal{T}\varphi_{H,F}(e)=\varphi_{\hat{H},\hat{F}}(\mathcal{T}e).
\label{HFTdual map}
\ena
{\it Proof.}
The section of $E$ has the form (\ref{HF-twisted section}).
Thus, applying the rule (\ref{T-duality rule}), the T-dual of this section becomes
\bea
&\mathcal{T} \varphi_{H,F}(e) \nn
=&(X_1 -a_i (X_1)dy )+fdy +(\xi_1 +i_{X_1}B_{2i} +fB_{1i})+(h- B_{1i} (X_1)) (\p_y+a_i) \\
=&(X_1-B_{1i}(X_1)\p_y) +h\p_y +(\xi_1 +i_{X_1}(B_{2i} -B_{1i}\wedge a_i)+ha_i )+(f- a_i (X_1)) (dy+B_{1i}).
\nonumber
\ena
On the other hand, the section of $\hat{E}$ has the form (\ref{HF-twisted section})
with replacing $(H,F)$ to $(\hat H,\hat F)$.
Thus, we have for $\mathcal{T}e=X_1 +h\p_y +\xi_1 +f dy$,
\bea
\varphi_{\hat H,\hat F}(\mathcal{T}e)
&=(X_1 -\hat{a}_i (X_1)\p_y )+h\p_y +(\xi_1 +i_{X_1}\hat{B}_{2i} +h\hat{B}_{1i} )
+(f- \hat{B}_{1i} (X_1)) (dy+\hat{a}_i) .
\ena
By comparing them, (\ref{HFTdual map}) holds by the identification
\bea
\hat{B}_{2i}=B_{2i} -B_{1i}\wedge a_i,\quad
\hat{B}_{1i}=a_i,\quad
\hat{a}_i=B_{1i}.
\ena
As a result, we have on one hand,
\bea
\hat{a}=dy+\hat{a}_i =dy+B_{1i}
\ena
so that the dual $F$-flux is written as 
\bea
\hat{F}=d\hat{a}=dB_{1i}=H_2.
\ena
On the other hand, the dual local $2$-form becomes 
\bea
\hat{B}_i
&=\hat{B}_{2i} -\hat{B}_{1i} \wedge \hat{a}\nn
&=(B_{2i} -B_{1i}\wedge a_i) -a_i \wedge (dy+B_{1i}) \nn
&= B_{2i} -a_i \wedge dy,
\ena
and correspondingly, it yields the dual $H$-flux as 
\bea
\hat{H}
&=dB_{2i} -da_i \wedge dy \nn
&=(dB_{2i} +da_i \wedge B_{1i})-da_i \wedge (dy +B_{1i}) \nn
&=(dB_{2i} +B_{1i} \wedge F) - F \wedge \hat{a} \nn
&=H_3 - F\wedge \hat{a}.
\ena
These results are summarized as
\bea
(H_3,H_2,F)\to (H_3,F,H_2).
\ena
{\it (End of the proof.)}

It is now straightforward to show that $\mathcal{T}$ is a morphism of Courant algebroids 
$\mathcal{T}: E \to \hat{E}$, or equivalently, a morphism of
$TN\oplus \langle \p_y\rangle \oplus T^*N \oplus \langle dy\rangle$ 
with the change of bracket from $[\cdot,\cdot]_{H,F}$ to $[\cdot,\cdot]_{\hat{H},\hat{F}}$.
\bea
&\langle \mathcal{T}e_1,\mathcal{T}e_2 \rangle =\langle e_1,e_2 \rangle , ~~
\rho|_{TN} (\mathcal{T}e)=\rho|_{TN} (e),\nn
&[\mathcal{T}e_1,\mathcal{T}e_2]_{\hat{H},\hat{F}} =\mathcal{T}[e_1,e_2]_{H,F}.
\ena
{\it Proof.}
Since the first two equations are obvious to hold, we show the third equation.
We know already that $[\mathcal{T}e_1,\mathcal{T}e_2]_C
=\mathcal{T}[e_1,e_2]_C$ in (\ref{T preserves Coutant}) for the Courant bracket.
By using (\ref{T preserves Coutant}), the definition of the twisted brackets (\ref{HF-twisted bracket}) and the key relation (\ref{HFTdual map}), 
we have
\bea
[\mathcal{T}e_1,\mathcal{T}e_2]_{\hat{H},\hat{F}}
&=\varphi_{\hat{H},\hat{F}}^{-1}
[\varphi_{\hat{H},\hat{F}}(\mathcal{T}e_1), \varphi_{\hat{H},\hat{F}}(\mathcal{T}e_2)]_C \nn
&=\varphi_{\hat{H},\hat{F}}^{-1}
[\mathcal{T}\varphi_{H,F}(e_1), \mathcal{T}\varphi_{H,F} (e_2)]_C \nn
&=\varphi_{\hat{H},\hat{F}}^{-1}\mathcal{T}[\varphi_{H,F}(e_1), \varphi_{H,F} (e_2)]_C \nn
&=\mathcal{T} \varphi_{H,F}^{-1} [\varphi_{H,F}(e_1), \varphi_{H,F} (e_2)]_C \nn
&=\mathcal{T}[e_1,e_2]_{H,F},
\ena
where we used $\varphi_{\hat{H},\hat{F}}^{-1}\mathcal{T}
 =\mathcal{T} \varphi_{H,F}^{-1}$.
{\it (End of the proof.)}\\
This shows the advantages of representing various twists in terms of bundle maps $\varphi$.


\section{Topological T-duality of Poisson-Generalized geometry}
Along the line of the above reformulation of the topological T-duality
in the generalized geometry,
here we will formulate the topological T-duality in the Poisson-generalized geometry,
and show the following aspects:
\begin{enumerate}

\item[{1)}] A Lie algebroid $(T^*N)_\theta \oplus {\mathbb R}$ appears in 
the Courant algebroid $(TM)_0\oplus (T^*M)_\theta $ in two different ways as
$(T^*N)_\theta \oplus \langle dy \rangle $ and $(T^*N)_\theta \oplus \langle \p_y \rangle $.
The topological T-duality without flux is formulated as an exchange of these two Lie algebroids.

\item[2)] A $Q$-flux is introduced associating with a twisting of the Lie algebroid 
$(T^*N)_\theta \oplus {\mathbb R}$.

\item[3)] In $(TM)_0\oplus (T^*M)_\theta $, 
the twist of 2) corresponds to two different twistings, $R_2$- and $Q$-twisting.
The former is the bivector part of the $R$-flux, while the latter is the $Q$-flux.

\item[4)] The topological T-duality exchanges $R_2$ and $Q$.
\end{enumerate}
To this end, let us first formulate the dimensional reduction of the Lie algebroid $(T^*M)_\theta$.

\subsection{Topological T-duality without flux} 

Consider a Poisson manifold $(M,\theta)$ with a Poisson structure $\theta$ and the Lie algebroid 
$(T^*M)_\theta$ of this Poisson manifold. 
As in the previous section, we 
consider that $M$ is a trivial $S^1$ bundle, 
$M=N\times S^1$ with local coordinates $(x^m,y)$.
Then a basic section $\xi$ of $(T^*M)_\theta$ has the form
\bea
\xi=\xi_1 + f dy, 
\ena
where $\xi_1 =\xi_m(x) dx^m\in \Gamma(T^*N)$ is a $1$-form on $N$, 
and $f(x) \in C^\infty (N)$.

The Poisson bivector $\theta$ on $M$ is also assumed to be basic, 
decomposed in general as
\bea
\theta =\theta_2+\theta_1 \wedge \p_y ,\quad
\theta_2=\frac{1}{2}\theta^{mn}(x)\p_m\wedge\p_n,\quad
\theta_1=\theta^{m}(x) \p_m.
\label{Poisson tensor}
\ena
where $\theta_2 \in \Gamma( \wedge^2 TN)$ is a bivector 
and $\theta_1 \in \Gamma(TN)$ is a vector field on $N$.
The condition $[\theta,\theta]_S=0$ for $\theta$ to be a Poisson bivector is equivalent to 
\bea
[\theta_2, \theta_2]_S=0, \quad [\theta_2, \theta_1]_S =0,
\ena
with respect to the Schouten bracket for $\Gamma(\wedge^\bullet TN)$.
That is, $\theta_2$ is a Poisson structure on $N$, 
while $\theta_1$ is a Poisson vector field\footnote{
By definition, $\theta_1= - \theta(dy)$ is a Hamiltonian vector field of the function $y$.} 
which preserves $\theta_2$.

We further assume that $\theta_1=0$ (equivalently $\theta (dy)=0$) in (\ref{Poisson tensor}),
that is, $\theta =\theta_2$ is a Poisson structure on the base space $N$.
In this case, we may identify $(T^*M)_{\theta, {\rm basic}}=(T^*N)_\theta \oplus \langle dy \rangle$ 
as vector bundles over $N$.
Then, the Koszul bracket between basic $1$-forms reduces to
\bea
[\xi_1+fdy , \eta_1+gdy]_\theta 
=[\xi_1, \eta_1]_{\theta} + \left({\cal L}_{\xi_1} g-{\cal L}_{\eta_1}f\right)dy,
\label{reduced Koszul bracket}
\ena
which is a counterpart of \eqref{KK-reduced TM}
See appendix \ref{app:reduction}, for a proof.
The image of the anchor map $\theta:{(T^*M)_\theta} \to TM$ is also restricted to $TN$, 
since $\theta (\xi_1 +fdy)=\theta (\xi_1)$ for $\theta_1=0$.

It is apparent that there is the same kind of structure as seen 
in \S 3.1 in the dimensional reduction of $TM$.
First, the $S^1$-reduced Koszul bracket (\ref{reduced Koszul bracket}) is 
the same as that of the Lie algebroid $A=(T^*N)_\theta  \oplus {\mathbb R}$ over $N$,
\bea
[\xi_1+f , \eta_1+g]_A
=[\xi_1, \eta_1]_{\theta} + \left({\cal L}_{\xi_1} g-{\cal L}_{\eta_1}f\right),
\label{bracket of A theta}
\ena
which is an analogue of \eqref{bracket of TN+R}.
The anchor maps are also identical because 
$\rc{\rho}_A(\xi_1+f )=\theta (\xi_1)=\rc{\rho}_{(T^*M)_\theta}(\xi_1 +fdy)$. 
Second, the Dirac structure 
$L={\rm span} \{dx^m, \p_y \}$ of the Courant algebroid $(TM)_0\oplus (T^*M)_\theta$ 
has the same bracket, 
corresponding to a counterpart of \eqref{CourantBasicdy},
\bea
[\xi_1+f\p_y , \eta_1+g\p_y]
=[\xi_1, \eta_1]_{\theta} + \left({\cal L}_{\xi_1} g-{\cal L}_{\eta_1}f\right)\p_y,
\ena
when restricting to basic sections
(see appendix \ref{app:reduction}, for a proof.).
That is, $(L)_{\rm basic}=(T^*N)_\theta \oplus \langle \p_y \rangle$.

Then, there is an isomorphism $\mathcal{T}: (T^*N)_\theta \oplus \langle dy \rangle\to 
(T^*N)_\theta \oplus \langle \p_y \rangle$ of these Lie algebroids, and 
it is extended to the automorphism of 
$(TN)_0 \oplus \langle \p_y \rangle \oplus (T^*N)_\theta \oplus \langle dy \rangle$ defined by 
the exchange of $\p_y$ and $dy$ as
\bea
\mathcal{T}: X_1 +f\p_y +\xi_1 + h dy \mapsto X_1 +h\p_y +\xi_1 + f dy.
\ena
This is the analogue of the topological T-duality for the Poisson-generalized geometry, 
in the case of vanishing fluxes.
In fact, we have 
\bea
&\langle  \mathcal{T}e_1,  \mathcal{T}e_2 \rangle= \langle  e_1, e_2\rangle , 
~~\rho ( \mathcal{T}e) =\rho (e), \nn
&[\mathcal{T}e_1, \mathcal{T}e_2]= \mathcal{T}[e_1, e_2], \label{T preserves Sasa}
\ena
where $[\cdot,\cdot]$ denotes the bracket of 
$(TN)_0 \oplus \langle \p_y \rangle \oplus (T^*N)_\theta \oplus \langle dy \rangle$,
which is also the $S^1$-reduced bracket of $(TM)_0\oplus (T^*M)_\theta$, given by
\bea
&[X_1 +f\p_y +\xi_1 + h dy, Y_1 +g\p_y +\eta_1 + k dy]\nn
=&[\xi_1, \eta_1]_{\theta} + \left({\cal L}_{\xi_1} k-{\cal L}_{\eta_1}h\right)dy\nn
+&{\cal L}_{\xi_1}Y_1 -{\cal L}_{\eta_1}X_1 
-\textstyle{\frac{1}{2}}d_\theta (i_{\xi_1}Y_1-i_{\eta_1}X_1)
+\textstyle{\frac{1}{2}}(d_\theta h g -hd_\theta g -d_\theta k f +kd_\theta f)\nn
+&\left({\cal L}_{\xi_1} g-{\cal L}_{\eta_1}f\right)\p_y.
\label{reduced Sasa bracket}
\ena
{\it Proof.} 
The first two equations in (\ref{T preserves Sasa}) are obvious to hold.
For the last equation, we need to show that (\ref{reduced Sasa bracket}) is valid.
Note first that
\bea
[\xi_1 + h dy, Y_1 +g\p_y ]
&={\cal L}_{\xi_1 + h dy}(Y_1 +g\p_y )-\textstyle{\frac{1}{2}}d_\theta (i_{\xi_1 + h dy}(Y_1 +g\p_y ))\nn
&={\cal L}_{\xi_1}Y_1 -\textstyle{\frac{1}{2}}d_\theta (i_{\xi_1}Y_1 )+ ({\cal L}_{\xi_1}g)\p_y 
+\textstyle{\frac{1}{2}}(d_\theta h g -hd_\theta g),
\ena
where 
\bea
{\cal L}_{\xi_1 + h dy}(Y_1 +g\p_y )
&=(i_{\xi_1 + h dy}d_\theta +d_\theta i_{\xi_1 + h dy})(Y_1 +g\p_y )\nn
&=i_{\xi_1 + h dy}(d_\theta Y_1 +(d_\theta g)\p_y ) +d_\theta (i_{\xi_1}Y_1 + hg )\nn
&={\cal L}_{\xi_1}Y_1 + ({\cal L}_{\xi_1}g)\p_y +(d_\theta h)g ,
\ena
is used.
From this and (\ref{reduced Koszul bracket}), we have the reduced bracket (\ref{reduced Sasa bracket}).
Then, it is straightforward to show the last equation in (\ref{T preserves Sasa}).
{\it (End of the proof.)}

The situation is summarized schematically as the following diagram:
\begin{align}
\xymatrix{
& (TM)_0\oplus (T^*M)_\theta \ar[ld]\ar[rd] & \\
(T^*N)_\theta \oplus \langle dy \rangle  \ar[rr]^\simeq
&& (T^*N)_\theta \oplus \langle \p_y \rangle .
}
\end{align}

We close this part with a remark on our two assumptions. 
We have assumed that sections are basic, and that $\theta$ is basic and
$\theta_1=0$ in (\ref{Poisson tensor}).
In the standard generalized geometry $TM\oplus T^*M$, 
the dimensional reduction is directly related to the $S^1$-invariance:
A basic vector field $X_1+f\p_y$ and a basic $1$-form $\xi_1+hdy$ is $S^1$-invariant, that is, 
invariant under the shift along the fiber direction generated by the vector field ${\p_y}$:
\bea
{\cal L}_{\p_y}(X_1+f\p_y)=0, \quad {\cal L}_{\p_y}(\xi_1+hdy)=0.
\ena
In the Poisson-generalized geometry $(TM)_0\oplus (T^*M)_\theta$, 
the reasoning by using the vector field ${\p_y}$ seems to be subtle, since in general a shift is 
generated by a $1$-form.
Nevertheless, this assumption is natural in the dimensional reduction scheme.
For example, for a surjection $p:M\to N$, a bundle $T^*N \oplus {\mathbb R}$ over $N$ has the pull-back 
$p^* T^*N$ over $M$, whose sections are identified with basic sections,
i.e., $\Gamma(T^*M)_{\rm basic}=\Gamma(p^* (TN\oplus {\mathbb R}))$.
Thus, the former assumption corresponds is needed if the T-duality is formulated on the base space $N$.

On the other hand, the latter assumption on the Poisson tensor 
is used to reduce the bracket and the anchor map to that of 
the Lie algebroid $TN\oplus {\mathbb R}$.
We may think it also as the invariance under the shift generated by the $1$-form ${dy}$.
For generic sections of $(TM)_0\oplus (T^*M)_\theta$ and for a Poisson tensor of the form 
(\ref{Poisson tensor}) which is not necessary basic, we have
\bea
{\cal L}_{dy}X
&={\cal L}_{\theta (dy)}X +\theta (i_{X}d^2y) 
=-{\cal L}_{\theta_1}X ,\nn
{\cal L}_{dy}\xi
&={\cal L}_{\theta (dy)}\xi -i_{\theta (\xi)}d^2y 
=-{\cal L}_{\theta_1} \xi.
\ena
They vanish if $\theta_1=0$.
In other words, our assumptions mean
the invariance by both ${\cal L}_{\p_y}$ and ${\cal L}_{dy}$.
As shown in appendix \ref{app:reduction}, however, these are rather strong assumptions
in order to obtain two isomorphic Lie algebroids and satisfy the T-duality property.
In particular, the T-duality can also be formulated using Lie algebroids over $M$, 
without assuming the dimensional reduction (see the case i) in appendix \ref{app:reduction}).
This is peculiar to the Poisson geometry where the exterior derivative $d_\theta$ depends on $\theta$.
Although we do not consider this possibility in this paper, it is worth to investigate this case further.

\subsection{$Q$-twisting of the Lie algebroid $(T^*N)_\theta  \oplus {\mathbb R}$}

In this subsection,
we give a definition of $Q$-flux given in a parallel manner of $F$-flux as discussed in \S 3.2.
Here, we investigate the twisting of the Lie algebroid $(T^*N)_\theta  \oplus {\mathbb R}$, 
which we call $Q$-twisting.
The strategy is the same as the case of $F$-twisting in \S 3.2.

Given a good cover $\{U_i\}$ of $N$ and a trivialization $(Q,\alpha_i,\gamma_{ij})$ 
of a $d_\theta$-closed bivector $Q$ such that
\bea
Q|_{U_i}=d_\theta\alpha_i, \quad 
\alpha_j -\alpha_i |_{U_{ij}}=d_\theta\gamma_{ij},
\ena 
we define a transition function $G_{ij}:U_{ij}\to GL(d)$ ($d=\dim N+1 =\dim M$) by
\bea
G_{ij}=\mat{1,d_\theta\gamma_{ij},0,1},
\ena
Then, since $G_{ij}$ satisfies the cocycle condition, we obtain a vector bundle over $N$
\bea
A=\coprod_{x\in N} (T^*U_i)_\theta \oplus {\mathbb R}_i /\sim.
\ena
Since each $(T^*U_i)_\theta \oplus {\mathbb R}_i \to U_i$ is a Lie algebroid and 
the gluing condition preserves the anchor map and the Lie bracket, $A$ is in fact a Lie algebroid,
which satisfies the exact sequence
\bea
0\to {\mathbb R} \to A \to (T^*N)_\theta \to 0.
\ena
The set $\{\alpha_i\}$ of local vector fields gives a splitting $\tilde{s}:(T^*N)_\theta\to A$, locally defined by
\bea
\tilde{s}(\xi_1)=\xi_1-\alpha_i (\xi_1)
\ena
for $\xi_1 \in (T^*U_i)_\theta$.
Since $\tilde{s}((T^*N)_\theta)$ is globally well-defined, we may identify 
$A\simeq \tilde{s}((T^*N)_\theta)\oplus {\mathbb R}$, and any section of $A$ is uniquely specified by
\bea
\tilde{s}(\xi_1)+f 
\ena
for $\xi_1 \in \Gamma ((T^*N)_\theta)$ and $f \in C^\infty (N)$.
The Lie bracket of these sections, which is  the counterpart of \eqref{Liebracket318}, is given by
\bea
[\tilde{s}(\xi_1)+f , \tilde{s}(\eta_1)+g ]_A 
=\tilde{s} ([\xi_1, \eta_1]_\theta )+{\cal L}_{\xi_1}g-{\cal L}_{\eta_1}f 
+i_{\xi_1}i_{\eta_1}Q,
\ena
which corresponds to the $Q$-twisted Lie bracket on $(T^*N)_\theta \oplus {\mathbb R}$:
\bea
[\xi_1+f , \eta_1+g ]_Q 
=[\xi_1, \eta_1]_\theta +{\cal L}_{\xi_1}g-{\cal L}_{\eta_1}f +i_{\xi_1}i_{\eta_1}Q.
\label{gauge Q-bracket}
\ena
{\it Proof.}
\bea
[\tilde{s}(\xi_1)+f , \tilde{s}(\eta_1)+g ]_A 
&=[\xi_1 +(f-\alpha_i (\xi_1)) , \eta_1 +(g-\alpha_i (\eta_1)) ]_{(T^*N)_\theta\oplus {\mathbb R}} \nn
&=[\xi_1,\eta_1]_\theta +{\cal L}_{\xi_1}(g-\alpha_i (\eta_1)) -{\cal L}_{\eta_1}(f-\alpha_i (\xi_1)) \nn
&=[\xi_1,\eta_1]_\theta +{\cal L}_{\xi_1}g-{\cal L}_{\eta_1}f
-{\cal L}_{\xi_1}(\alpha_i (\eta_1)) +{\cal L}_{\eta_1}(\alpha_i (\xi_1)) \nn
&=[\xi_1,\eta_1]_\theta -\alpha_i ([\xi_1,\eta_1]_\theta) +{\cal L}_{\xi_1}g-{\cal L}_{\eta_1}f
+ i_{\xi_1} i_{\eta_1}d_\theta \alpha_i \nn
&=\tilde{s}([\xi_1,\eta_1]_\theta) +{\cal L}_{\xi_1}g-{\cal L}_{\eta_1}f
+ i_{\xi_1} i_{\eta_1}d_\theta \alpha_i,
\ena
where
\bea
-{\cal L}_{\xi_1}(\alpha_i (\eta_1)) +{\cal L}_{\eta_1}(\alpha_i (\xi_1))
&=-d_\theta i_{\xi_1} i_{\eta_1}\alpha_i 
-i_{\xi_1}d_\theta i_{\eta_1}\alpha_i +{\cal L}_{\eta_1}i_{\xi_1}\alpha_i  \nn
&=0-i_{\xi_1}({\cal L}_{\eta_1} -i_{\eta_1}d_\theta )\alpha_i +{\cal L}_{\eta_1}i_{\xi_1}\alpha_i  \nn
&=[{\cal L}_{\eta_1},i_{\xi_1}]\alpha_i + i_{\xi_1} i_{\eta_1}d_\theta \alpha_i \nn
&=-i_{[\xi_1,\eta_1]_\theta}\alpha_i + i_{\xi_1} i_{\eta_1}d_\theta \alpha_i ,
\ena
is used.{\it (End of the proof)}\\

By construction, $Q$ is a 
$d_\theta$-closed bivector $Q \in \Gamma(\wedge^2 TN)$ on $N$, and 
is regarded as the field strength of the gauge potential $\{\alpha_i\}$.
In local coordinates, $Q$ is written as
\bea
Q|_{U_i}&=d_\theta \alpha_i 
=[\theta, \alpha_i]_S \nn
&=\textstyle{\frac{1}{2}}\left(\theta^{ml}\p_l \alpha_i^{n} -\theta^{nl}\p_l \alpha_i^{m}
-\alpha_i^{l}\p_l \theta^{mn}\right) \p_m \wedge \p_n.
\ena
Although it has a rather complicated form, $Q$ is an abelain-type field strength.
In fact, there is a gauge symmetry of the form
\bea
Q \mapsto  Q, \quad
\alpha_i \mapsto \alpha_i + d_\theta f_i, \quad \gamma_{ij}\mapsto \gamma_{ij}+f_j -f_i,
\ena
for a set of local gauge parameters $f_i \in C^\infty(U_i)$,
and this transformation preserves $Q$.
On the other hand, a shift of $\{\alpha_i\}$ 
\bea
\alpha_i \mapsto  \alpha_i +\alpha
\ena
by a global vector field $\alpha \in \Gamma(TN)$, corresponds to a change of splitting.
This changes $Q \mapsto Q+d_\theta \alpha$ but preserves its cohomology class $[Q]\in H^2_\theta (N)$
in the Poisson cohomology.

Recall that the $F$-twisting of the Lie algebroid $TN\oplus {\mathbb R}$,
leads to $A=\tilde{s}(TN)\oplus {\mathbb R}$, but when considering it as a tangent bundle of some space, 
it is identified with $A=TP/U(1)$.
That is, the underlying topological space has been changed from $M=N\times S^1$ 
to $P$.
In this sense, the Lie algebroid $A=\tilde{s}((T^*N)_\theta)\oplus {\mathbb R}$ 
here would be identified with some Poisson-version of a principal bundle,
but we do not know what this is.
It would be related to a question what a non-geometric space is,
and in general it is a challenging issue.
This is why we need a reformulation of the topological T-duality so far, 
where a manifold $M$ is unchanged and
without recourse to the use of ``principal bundles".

\subsection{$R$ and $Q$-fluxes in Poisson-generalized geometry}

Similar to the case of $F$-twisting in the standard Courant algebroid,
the $Q$-twisted Lie algebroid described in \S 4.2 can appear in two different places
in $(TM)_0\oplus (T^*M)_\theta$.
The structure is completely analogous to the case of $(H,F)$-fluxes:
\begin{align}
\xymatrix{
& (TM)_0\oplus (T^*M)_\theta  \ar[ld]\ar[d]\ar[rd] & \\
(T^*N)_\theta \oplus \langle dy \rangle \ar@(ld,rd)_{Q} 
& (TN)_0\oplus (T^*N)_\theta  \ar@(ld,rd)_{R_3}
& (T^*N)_\theta \oplus \langle \p_y \rangle \ar@(ld,rd)_{R_2},
}
\end{align}
Here $Q$ denotes the $Q$-twisting of $(T^*N)_\theta\oplus \langle dy\rangle$,
while $R_2$ denotes the $Q$-twisting of $(T^*N)_\theta \oplus \langle \p_y \rangle$.
As we will elaborate on, we call the former as a $Q$-flux, the counterpart of a $F$-flux,
because it is glued by $\beta$-diffeomorphism.
On the other hand, the latter is a bivector part of the $R$-flux, 
since it is a result of local $\beta$-transformation.
By combining with $R_3$, appearing in the $R$-twisting of $(T^*N)_\theta\oplus (TN)_0$,
we have an $R$-flux in the total space $M$ as
\bea
R=R_3-R_2\wedge \alpha ,
\label{R decomposition}
\ena
where $\alpha$ is defined locally by\footnote{{It is an $A$-$1$ form on $A$ and is regarded as a global vector field on the ``principal bundle".}}
\bea
\alpha|_{U_i} =\p_y +\alpha_i
\label{alpha}
\ena
such that $Q=d_\theta \alpha$.

In the following, we describe this structure in more detail.
We first twist $(T^*N)_\theta\oplus \langle dy\rangle$ by $Q$ to obtain a Courant algebroid 
$A\oplus A^*$ with $A=\tilde{s}((T^*N)_\theta) \oplus \langle dy\rangle$, and then we
twist $A\oplus A^*$ by $R$ to obtain a Courant algebroid $E=s(A)\oplus A^*$.

\subsubsection{$Q$-flux}

Given data $(Q, \alpha_i,\gamma_{ij})$ in \S 4.2, we obtain a $Q$-twisted Lie algebroid 
$A=\tilde{s}((T^*N)_\theta) \oplus \langle dy\rangle$, whose section has the form,
analogue of \eqref{323},
\bea
\tilde{s}(\xi_1 )+h dy, ~~\tilde{s}(\xi_1 )=\xi_1 -\alpha_i (\xi_1)dy.
\ena
As in \S 3.3, this $Q$-twisting also affects its dual $A^*$, reflecting in the decomposition
\bea
X_1 +f\alpha,
\ena
where $\alpha$ is given in (\ref{alpha}).
The following relations hold:
\bea
i_{\tilde{s}(\xi_1)}\alpha =0, ~~i_{dy} \alpha=1, ~~ 
i_{\tilde{s}(\xi_1)}X_1 =i_{\xi_1}X_1 , ~~i_{dy} X_1=0,
\ena
to be compared with \eqref{326}.
By combining them, the $Q$-twisting of the Courant algebroid $(TM)_0\oplus (T^*M)_\theta$
is specified by a bundle map 
$\varphi_Q: (TN)_0\oplus \langle \p_y\rangle \oplus (T^*N)_\theta\oplus \langle dy\rangle
\to A\oplus A^*$, locally given by
\bea
\varphi_Q  =\mat{e^{-\alpha_i \wedge dy},0,0,e^{-\alpha_i \wedge dy}}.
\label{Q map}
\ena
Indeed, it is shown that 
\bea
\varphi_Q (X_1 +f\p_y+\xi_1 +hdy)=X_1 + f\alpha +\tilde{s}(\xi_1)+hdy,
\ena
because of 
\bea
&e^{-\alpha_i \wedge dy}(X_1 +f\p_y )=X_1 +f\p_y +f \alpha_i =X_1 + f\alpha,\nn
&e^{-\alpha_i \wedge dy}(\xi_1 +hdy)=\xi_1 +hdy -\alpha_i (\xi_1) dy  =\tilde{s}(\xi_1)+hdy.
\ena

Correspondingly, the transition function is given by
\bea
G^Q_{ij} =\mat{e^{-d_\theta \lambda_{ij} \wedge dy},0,0,e^{-d_\theta \lambda_{ij} \wedge dy}}.
\ena 
It is a $\beta$-diffeomorphism $e^{{\cal L}_{\zeta_{ij}}}$ 
generated by the $1$-form $\zeta_{ij}=\lambda_{ij} dy \in \langle dy \rangle$ 
on $U_{ij}$.\\
{\it Proof.} We see that the action of ${\cal L}_{\zeta_{ij}}$ is equivalent to the gluing condition.
\bea
{\cal L}_{\zeta_{ij}}(X_1 +f\p_y )
&={\cal L}_{\theta (\zeta_{ij})}(X_1 +f\p_y )+\theta (i_{X_1 +f\p_y }d\zeta_{ij})
=\theta (i_{X_1 +f\p_y }d\lambda_{ij}\wedge dy ) \nn
&=\theta (d\lambda_{ij} (X_1) dy - f\lambda_{ij}) 
=-f d_\theta \lambda_{ij},\nn
{\cal L}_{\zeta_{ij}}(\xi_1 +hdy)
&={\cal L}_{\theta (\zeta_{ij})}(\xi_1 +hdy) -i_{\theta (\xi_1 +hdy)}d\zeta_{ij}
=-i_{\theta (\xi_1)}(d\lambda_{ij} \wedge dy) \nn
&=-\theta (\xi_1,d\lambda_{ij})dy =\theta (d\lambda_{ij},\xi_1)dy 
=-(d_\theta \lambda_{ij}) (\xi_1) dy, 
\ena
where $\theta (dy)=0$ is used.
{\it (End of the proof.)}\\
Because ${\cal L}_{\zeta_{ij}}\theta =d_\theta i_{\zeta_{ij}} \theta =0$, 
this $\beta$-diffeomorphism satisfies the condition for the symmetry \cite{AMSW}.
This shows that $Q$-fluxes are associated with $\beta$-diffeomorphisms.

The $Q$-twisted bracket on 
$(TN)_0\oplus \langle \p_y\rangle \oplus (T^*N)_\theta\oplus \langle dy\rangle$ is defined 
through (\ref{Q map}) as
\bea
[e_1,e_2]_Q :=\varphi_Q^{-1}[\varphi_Q (e_1), \varphi_Q (e_2)].
\ena
The explicit form will be given by
\bea
&[e_1, e_2 ]_{Q}
=[e_1, e_2 ]
+(i_{\xi_1}i_{\eta_1}Q )dy  +gi_{\xi_1}Q -fi_{\eta_1}Q,
\label{explicit Q-bracket}
\ena
for $e_1=X_1+f\p_y +\xi_1 +hdy$ and $e_2=Y_1+g\p_y +\eta_1 +kdy$.\\
{\it Proof.}
We first calculate each term in 
\bea
[\varphi_Q (e_1), \varphi_Q (e_2)]
&=[X_1 + f\alpha +\tilde{s}(\xi_1)+hdy, Y_1 + g\alpha +\tilde{s}(\eta_1)+kdy]\nn
&=[\tilde{s}(\xi_1) +hdy, \tilde{s}(\eta_1) +kdy]_\theta 
+[\tilde{s}(\xi_1)+hdy, Y_1 + g\alpha ]
+[X_1 + f\alpha , \tilde{s}(\eta_1)+kdy].
\ena
Here the first term
\bea
[\tilde{s}(\xi_1) +hdy, \tilde{s}(\eta_1) +kdy]_\theta
&=\tilde{s}([\xi_1,\eta_1]_\theta) +({\cal L}_{\xi_1} k -{\cal L}_{\eta_1} h) dy
+(i_{\xi_1}i_{\eta_1} Q) dy \nn
&=\varphi_Q ([\xi_1 +hdy ,\eta_1 +kdy] +(i_{\xi_1}i_{\eta_1} Q) dy),
\ena
is a consequence of (\ref{gauge Q-bracket}).
The second term is written as
\bea
[\tilde{s}(\xi_1)+hdy, Y_1 + g\alpha ] 
&={\cal L}_{\tilde{s}(\xi_1) +hdy}( Y_1 + g\alpha) 
-\textstyle{\frac{1}{2}}d_\theta i_{\tilde{s}(\xi_1) +hdy}( Y_1 + g\alpha) \nn
&=i_{\tilde{s}(\xi_1) +hdy} d_\theta ( Y_1 + g\alpha)
+\textstyle{\frac{1}{2}}d_\theta i_{\tilde{s}(\xi_1) +hdy}( Y_1 + g\alpha) \nn
&=i_{\xi_1} d_\theta Y_1 + (i_{\xi_1} d_\theta g) \alpha -h d_\theta g +g  i_{\xi_1} Q 
+\textstyle{\frac{1}{2}}d_\theta (i_{\xi_1} Y_1 +h g) \nn
&={\cal L}_{\xi_1} Y_1 + ({\cal L}_{\xi_1}g) \alpha 
-\textstyle{\frac{1}{2}}d_\theta i_{\xi_1} Y_1 +\textstyle{\frac{1}{2}}(d_\theta h g-hd_\theta g)
+g  i_{\xi_1} Q \nn
&=\varphi_Q ([\xi_1+hdy, Y_1 + g\alpha ] +g  i_{\xi_1} Q),
\ena
and similar for the third term.
Therefore, it leads to
\bea
[\varphi_Q (e_1), \varphi_Q (e_2)]
&=\varphi_Q ([e_1,e_2]+(i_{\xi_1}i_{\eta_1} Q) dy +g  i_{\xi_1} Q -fi_{\eta_1} Q).
\ena
{\it (End of the proof.)}\\

By construction, our $Q$-flux proposed here in the Poisson-generalized geometry 
is completely analogous to the geometric $F$-flux in the standard generalized geometry.
However, there are of course some differences, 
since the underlying Lie algebroid $(T^*N)_\theta$ is different from $TN$.

Recall that an $F$-flux can be in general characterized either by the Lie bracket of frame vector fields 
$e_a$ $(a=1, \cdots, {\rm dim} M)$ or the Maurer-Cartan equation of $1$-forms $e^a$ as\footnote{
It is also consistent with the relation $[e_a,e_b]_C=f_{ab}^c e_c$ and $[e^c,e_a]_C =-f_{ab}^c e^b$ 
of the Courant bracket.}
\bea
[e_a,e_b]=f_{ab}^c e_c, ~~de^c =-\textstyle{\frac{1}{2}}f^c_{ab}e^a \wedge e^b.
\ena
In the case of the $F$-flux in \S 3.3.1, if we define locally
\bea
&e_m =\tilde{s}(\p_m)=\p_m -A_i (\p_m) \p_y, ~~ e_y=\p_y,\\
&e^m=dx^m, ~~e^y=A=dy+A_i ,
\ena
then it is easy to show that 
\bea
[e_m,e_n]= -F_{mn} e_y, ~~de^y =\textstyle{\frac{1}{2}}F_{mn}e^m \wedge e^n,
\ena
with $F_{mn}=F(\p_m,\p_n)$.
It says that $f_{mn}^y =-F_{mn}$ and zero for others.

On the other hand, for the $Q$-flux here, if we define fame fields as
\bea
&e_m =\p_m, ~~ e_y=\alpha =\p_y +\alpha_i,\\
&e^m=\tilde{s}(dx^m)=dx^m-\alpha_i (dx^m)dy, ~~e^y=dy ,
\ena
then it can be shown that
\bea
&d_\theta e_m =[\theta,\p_m]_S= -\p_m \theta^{nl} e_n\wedge e_l,\nn
&d_\theta e_y =d_\theta \alpha_i =\textstyle{\frac{1}{2}} Q^{nl} e_n\wedge e_l,\nn
&[e^m,e^n]=\tilde{s}([dx^m,dx^n]_\theta )- Q(dx^m,dx^n) dy
=\p_l \theta^{mn} e^l - Q^{mn} e^y,
\ena
with $Q^{mn}=Q(dx^m,dx^n)$.
It says that they satisfy the relations\footnote{
It is consistent with the relations
$[e^a,e^b]=q^{ab}_c e^c$ and $[e^a,e_c]=-q^{ab}_c e_b$
of the bracket in the Poisson-generalized geometry.}
\bea
[e^a,e^b]=q^{ab}_c e^c, ~~d_\theta e_c =-\textstyle{\frac{1}{2}}q_c^{ab}e_a \wedge e_b,
\ena
with $q^{mn}_l =\p_l \theta^{mn}$, $q^{mn}_y=-Q^{mn}$ and zero for others.
We observe that $q^{mn}_l$ coming from the structure of $(T^*N)_\theta$ does not vanish 
even when $Q=0$.

\subsubsection{$(R,Q)$-flux}
By twisting $A\oplus A^*$ with $R$ further, we obtain a Courant algebroid $E$, 
specified by a bundle map $\varphi_{R} : A \oplus A^* \to E$.
Hence,
the twisting of $(TN)_0\oplus \langle \p_y\rangle \oplus (T^*N)_\theta \oplus \langle dy\rangle$ 
by both $R$ and $Q$-flux is specified by the bundle map
\bea
\varphi_{R,Q} := \varphi_R \varphi_Q
&=\mat{1,\beta_i,0,1}\mat{e^{-\alpha_i \wedge dy},0,0,e^{-\alpha_i \wedge dy}}.
\label{phi RQ}
\ena
Here the set of local bivectors $\{\beta_i\}$ defines a splitting
$s: A \to E$, and $A$ carries already the information of the $Q$-flux.
Correspondingly, any basic $k$-vector on $M$ is decomposed with respect to $\alpha$ in (\ref{alpha}) into
\bea
V  =V_k +V_{k-1}\wedge \alpha,
\ena
where $k$ and $k-1$ are degrees as polyvectors on $N$.
In particular, we decompose the global $3$-vector $R$ as in (\ref{R decomposition}), and 
local bivector $B_i$ in (\ref{phi RQ}) into
\bea
&\beta_i=\beta_{2i} -\beta_{1i} \wedge \alpha, \quad \beta_{2i}
 \in \Gamma(\wedge^2 TU_i) ,~ \beta_{1i} \in \Gamma(TU_i).
\ena
Then, by noting 
\bea
R|_{U_i}&=d_\theta \beta_i =(d_\theta \beta_{2i}+\beta_{1i} \wedge Q)-d_\theta \beta_{1i} \wedge \alpha,
\ena
we should identify 
\bea
R_3|_{U_i}=d_\theta \beta_{2i}+\beta_{1i} \wedge Q,\quad
{R}_2|_{U_i}=d_\theta \beta_{1i}.
\ena
Thus, the local vector field $\beta_{1i}$ is the gauge potential for the $R_2$-flux.

With this decomposition, (\ref{phi RQ}) says that the section of $E$
is locally given by
\bea
\varphi_{R,Q}(e)
&=\varphi_{R}(X_1 +f\alpha +\tilde{s}(\xi_1 )+hdy) \nn
&=X_1 +f\alpha +\tilde{s}(\xi_1 )+hdy+ i_{\tilde{s}(\xi_1 )+hdy} \beta_i \nn
&=X_1 +f\alpha +\tilde{s}(\xi_1 )+hdy+ i_{\tilde{s}(\xi_1 )+hdy} (\beta_{2i} -\beta_{1i} \wedge \alpha) \nn
&=(X_1 +i_{\xi_1}\beta_{2i} + h\beta_{1i}) +(f-\beta_{1i}(\xi_1 ))\alpha +\tilde{s}(\xi_1 )+hdy ,
\label{RQ-twisted section}
\ena
for $e=X_1+f\p_y +\xi_1 +hdy$.

The $(R,Q)$-twisted bracket on 
$(TN)_0\oplus \langle \p_y\rangle \oplus (T^*N)_\theta \oplus \langle dy\rangle$ is defined by 
\bea
[e_1,e_2]_{R,Q} :=\varphi_{R,Q}^{-1}[\varphi_{R,Q}(e_1), \varphi_{R,Q}(e_2)].
\label{RQ-twisted bracket}
\ena
By a direct calculation using (\ref{RQ-twisted section}), we have explicitly
\bea
&[e_1, e_2 ]_{R,Q}\nn
&=[e_1, e_2 ]
-i_{\xi_1}i_{\eta_1}R_3
+(i_{\xi_1}i_{\eta_1}Q )dy +\left(i_{\xi_1}i_{\eta_1}R_2\right)\alpha 
+gi_{\xi_1}Q -fi_{\eta_1}Q -hi_{\eta_1}R_2 +ki_{\xi_1}R_2,
\ena
for $e_1=X_1+f\p_y +\xi_1 +hdy$ and $e_2=Y_1+g\p_y +\eta_1 +kdy$.\\
{\it Proof.} 
We first show the validity of the $R$-twisted bracket.
For $e_1=X+\xi$ and $e_2=Y+\eta$, we calculate
\bea
[\varphi_{R}(e_1), \varphi_{R}(e_2)]
&=[X+i_{\xi}\beta_i +\xi, Y+ i_{\eta}\beta_i +\eta ] \nn
&=[X+\xi, Y+\eta ]+ [\xi, i_{\eta}\beta_i] + [i_{\xi}\beta_i, \eta ]\nn
&=[X+\xi, Y+\eta ] + \beta_i ([\xi,\eta]_\theta ) -i_\xi i_\eta R \nn
&=\varphi_{R}([X+\xi, Y+\eta ] -i_\xi i_\eta R), 
\ena
where 
\bea
[\xi, i_{\eta}\beta_i]+ [i_{\xi}\beta_i, \eta]
&={\cal L}_{\xi}i_{\eta}\beta_i -{\cal L}_{\eta}i_{\xi}\beta_i 
-\textstyle{\frac{1}{2}}d_\theta (i_{\xi}i_{\eta}\beta_i -i_{\eta}i_{\xi}\beta_i)\nn
&=i_{\xi}d_\theta i_{\eta}\beta_i -{\cal L}_{\eta}i_{\xi}\beta_i \nn
&=i_{\xi}({\cal L}_{\eta} -i_\eta d_\theta )\beta_i -{\cal L}_{\eta}i_{\xi}\beta_i \nn
&=-[{\cal L}_{\eta}, i_{\xi}]\beta_i -i_{\xi}i_\eta R \nn
&=i_{[\xi,\eta]_\theta } \beta_i -i_{\xi}i_\eta R
\ena
is used.
This shows $[X+\xi, Y+\eta ]_R=[X+\xi, Y+\eta ]-i_\xi i_\eta R$, which is also valid for 
the decomposition like $e_1=X_1+f\alpha +\tilde{s}(\xi_1) +hdy$.
Thus, 
\bea
i_{\xi}i_{\eta}R
&=i_{\tilde{s}(\xi_1) +hdy} i_{\tilde{s}(\eta_1) +kdy} (R_3 -R_2 \wedge \alpha) \nn
&=i_{\tilde{s}(\xi_1) +hdy} (i_{\eta_1}R_3 -i_{\eta_1}R_2 \wedge \alpha -k R_2) \nn
&=i_{\xi_1}i_{\eta_1}R_3 -(i_{\xi_1}i_{\eta_1}R_2) \alpha +h i_{\eta_1}R_2 -k i_{\xi_1} R_2.
\ena
By combining with the explicit form of the $Q$-twisted bracket (\ref{explicit Q-bracket}), 
we obtain the result.\\
{\it (End of the proof.)}

\subsection{Topological T-duality with $R$ and $Q$-fluxes}

The topological T-duality is a bundle map of $S^1$-reduced Courant algebroid
$(TN)_0\oplus \langle \p_y\rangle \oplus (T^*N)_\theta \oplus \langle dy\rangle$ acting as
\bea
\mathcal{T}:X_1 +f \p_y +\xi_1 +h dy \mapsto X_1 +h\p_y +\xi_1 +f dy.
\label{T-duality rule2}
\ena
After $(R,Q)$-twisting, it is also regarded as a map 
$\mathcal{T}:E\to \hat{E}$ of twisted Courant algebroids.
The structure is exactly the same as the case of the $(H,F)$-twisting in \S 3.4.
The key relation between sections of $E$ and $\hat{E}$ in this case, similar to (\ref{HFTdual map}), is
\bea
\mathcal{T}\varphi_{R,Q}(e)=\varphi_{\hat{R},\hat{Q}}(\mathcal{T}e),
\label{RQTdual map}
\ena
provided that
\bea
(R=R_3 -R_2 \wedge \alpha, Q) \mapsto (\hat{R}=R_3 -Q \wedge \hat{\alpha}, R_2).
\ena
{\it Proof.}
The section of $E$ has the form (\ref{RQ-twisted section}).
Thus, applying the rule (\ref{T-duality rule2}), the T-dual of this section becomes
\bea
\mathcal{T} \varphi_{R,Q}(e)
&=(X_1 +i_{\xi_1}\beta_{2i} + h\beta_{1i}) +(f-\beta_{1i}(\xi_1 ))(dy +\alpha_i) 
+(\xi_1 -\alpha_i (\xi_1) \p_y)+h\p_y \\
\nonumber
&=(X_1 +i_{\xi_1}(\beta_{2i} -\beta_{1i} \wedge \alpha_i )+ f\alpha_i ) 
+(h-\alpha_i (\xi_1)) (\p_y + \beta_{1i})  
+(\xi_1 -\beta_{1i}(\xi_1 )dy)  +f dy.
\ena
On the other hand, the section of $\hat{E}$ has the form (\ref{RQ-twisted section})
with replacing $(R,Q)$ to $(\hat R,\hat Q)$.
Thus, we have for $\mathcal{T}e=X_1 +h\p_y +\xi_1 +f dy$,
\bea
\varphi_{\hat R,\hat Q}(\mathcal{T}e)
&=(X_1 +i_{\xi_1}\hat \beta_{2i} + f\hat \beta_{1i}) 
+(h-\hat \beta_{1i}(\xi_1 ))(\p_y + \hat \alpha_i) +(\xi_1 -\hat \alpha_i (\xi_1) dy)+fdy .
\ena
By comparing them, (\ref{RQTdual map}) holds by the identification
\bea
\hat{\beta}_{2i}=\beta_{2i} -\beta_{1i}\wedge \alpha_i,\quad
\hat{\beta}_{1i}=\alpha_i,\quad
\hat{\alpha}_i=\beta_{1i}.
\ena
As a result, we have on one hand,
\bea
\hat{\alpha}=\p_y+\hat{\alpha}_i =\p_y+\beta_{1i},
\ena
so that the dual $Q$-flux is written as 
\bea
\hat{Q}=d_\theta \hat{\alpha}=d_\theta \beta_{1i}=R_2.
\ena
On the other hand, the dual local bivector becomes 
\bea
\hat{\beta}_i
&=\hat{\beta}_{2i} -\hat{\beta}_{1i} \wedge \hat{\alpha}\nn
&=(\beta_{2i} -\beta_{1i}\wedge \alpha_i) -\alpha_i \wedge (\p_y+\beta_{1i}) \nn
&= \beta_{2i} -\alpha_i \wedge \p_y,
\ena
and correspondingly, it yields the dual $R$-flux as
\bea
\hat{R}
&=d_\theta \beta_{2i} -d_\theta \alpha_i \wedge \p_y \nn
&=(d_\theta \beta_{2i} +d_\theta \alpha_i \wedge \beta_{1i})-d_\theta \alpha_i \wedge (\p_y +\beta_{1i}) \nn
&=(d_\theta \beta_{2i} +\beta_{1i} \wedge Q)-Q \wedge \hat \alpha \nn
&=R_3 -Q \wedge \hat \alpha.
\ena
These results are summarized as
\bea
(R_3,R_2,Q)\to (R_3,Q,R_2).
\ena
{\it (End of the proof.)}

It is now straightforward to show that $\mathcal{T}$ is a morphism of Courant algebroids.
\bea
&\langle \mathcal{T}e_1,\mathcal{T}e_2 \rangle =\langle e_1,e_2 \rangle , ~~
\rho (\mathcal{T}e)=\rho (e),\nn
&[\mathcal{T}e_1,\mathcal{T}e_2]_{\hat{R},\hat{Q}} =\mathcal{T}[e_1,e_2]_{R,Q}.
\ena
{\it Proof.}
By using (\ref{T preserves Sasa}), (\ref{RQ-twisted bracket}) and (\ref{RQTdual map}), 
we have
\bea
[\mathcal{T}e_1,\mathcal{T}e_2]_{\hat{R},\hat{Q}}
&=\varphi_{\hat{R},\hat{Q}}^{-1}
[\varphi_{\hat{R},\hat{Q}}(\mathcal{T}e_1), \varphi_{\hat{R},\hat{Q}}(\mathcal{T}e_2)] \nn
&=\varphi_{\hat{R},\hat{Q}}^{-1}
 [\mathcal{T}\varphi_{R,Q}(e_1), \mathcal{T}\varphi_{R,Q} (e_2)]\nn
&=\varphi_{\hat{R},\hat{Q}}^{-1}\mathcal{T}[\varphi_{R,Q}(e_1), \varphi_{R,Q} (e_2)] \nn
&=\mathcal{T} \varphi_{R,Q}^{-1} [\varphi_{R,Q}(e_1), \varphi_{R,Q} (e_2)] \nn
&=\mathcal{T}[e_1,e_2]_{R,Q}.
\ena
where we used $\varphi_{\hat{R},\hat{Q}}^{-1} \mathcal{T}= \mathcal{T} \varphi_{R,Q}^{-1}$.
{\it (End of the proof.)}

\section{Conclusion and Discussion}

In this paper we have studied the topological T-duality  
in the standard generalized geometry as well as the Poisson-generalized geometry, 
and examined the property of the fluxes related to the twisting of the underlying Courant algebroids.

By the $S^1$-dimensional reduction of the generalized tangent bundle 
$TM\oplus T^*M$, we reformulated the topological 
T-duality as an exchange of two isomorphic Lie algebroids.
By using this reformulation, 
we proved the topological T-duality in the Poisson-generalized geometry 
based on the Courant algebroid $(TM)_0\oplus (T^*M)_\theta$.

In the standard generalized geometry, when $TM\oplus T^*M$ is twisted, $H$-fluxes and geometric $F$-fluxes are exchanged under the topological T-duality.
In particular, a $F$-flux associated with diffeomorphism is obtained form the 
$2$-form part $H_2$ of a $H$-flux associated with $B$-gauge transformations.
Similarly, in the Poisson generalized geometry, as proposed in the previous paper \cite{AMSW},
an $R$-flux appears associated with the $\beta$-gauge transformation. 
Then applying the topological T-duality, as an natural analogue of the geometric $F$-flux, 
we obtain a bivector flux associated with $\beta$-diffeomorphisms. 
We proposed to identify it with a $Q$-flux, which corresponds to the definition of $Q$-fluxes 
as $Q$-twistings of the Lie algebroid $(T^*N)_\theta \oplus {\mathbb R}$.
As a result, we obtained a clear classification of four kinds of fluxes in terms of symmetries as 
\begin{center}
\begin{tabular}{ll}
$H$: & $B$-gauge transformation, \\
$F$: & diffeomorphism, \\
$R$: & $\beta$-gauge transformation,\\
$Q$: & $\beta$-diffeomorphism.
\end{tabular}
\end{center}
The  twistings corresponding to these fluxes are expressed in an unified way by introducing the bundle maps, 
which can be used also to formulate the twisted brackets.

Using this formalism, we showed that $(R_2,Q)$-fluxes are exchanged by 
the topological T-duality of the Poisson-generalized geometry,
just as $(H_2,F)$-fluxes are exchanged by the topological T-duality of the standard generalized geometry.
However, it is apparent that the topological T-duality never relate 
$(H,F)$-fluxes and $(R,Q)$-fluxes in this way, 
because two underlying Courant algebroids are different.
In order to relate all the fluxes and complete the duality chain, 
we need another kind of map than the topological T-duality.

The paper includes a reformulation of the standard topological T-duality for the circle bundles. 
Recently, an extension to $SU(2)$-bundle, called the spherical T-duality, 
is proposed \cite{Bouwknegt:2014oka}.
Along the same line, our viewpoint using Lie algebroids would help to understand such cases further.

We also pointed out the possibility to formulate the T-duality without dimensional reduction.
As already stated, it is worth to investigate this case further.
In particular, it is interesting to apply it to field theories with keeping all the Kaluza-Klein modes.
To this end, we need to formulate a supergravity theory with $R$ and $Q$-fluxes
based on the Poisson-generalized geometry.
As stated already in the previous paper \cite{AMSW}, the relevance of this new geometry
in physics should be further studied in the context of string theory.
We will come to this point in the future publication.


\section*{Acknowledgments}
Authors would like to thank 
the members of the particle theory and cosmology group, 
in particular U.~Carow-Watamura for helpful comments and discussions.
H.~M. is supported by Tohoku University
Institute for International Advanced Research and Education.

 \appendix

\section{Proof of  \eqref{T preserves Coutant} \label{autoofT}}

The invariance of the inner product and anchor is apparent. 
To see the compatibility with the Courant bracket, we need the formula of the Courant bracket of the basic sections.
From (\ref{KK-reduced TM}) and 
\bea
[X_1+ f\p_y, \eta_1 +k dy]_C
&={\cal L}_{X_1+ f\p_y } (\eta_1 +k dy) 
-\frac{1}{2}d \left( i_{X_1+ f\p_y } (\eta_1 +k dy)\right)\nn
&={\cal L}_{X_1} \eta_1 -\frac{1}{2}d \left( i_{X_1} \eta_1 \right)
+{\cal L}_{X_1 +f\p_y} (kdy) -\frac{1}{2}d (fk)\nn
&={\cal L}_{X_1} \eta_1 -\frac{1}{2}d \left( i_{X_1} \eta_1 \right)
+({\cal L}_{X_1}k)dy +k{\cal L}_{f\p_y} (dy)-\frac{1}{2}d (fk)\nn
&={\cal L}_{X_1} \eta_1 -\frac{1}{2}d \left( i_{X_1} \eta_1 \right)
+({\cal L}_{X_1}k)dy +\frac{1}{2}(df k -fdk),
\ena
we have
\bea
&[X_1+ f\p_y +\xi_1 +h dy, Y_1+ g\p_y +\eta_1 +k dy]_C \nn
=&[X_1,Y_1] + \left({\cal L}_{X_1} g -{\cal L}_{Y_1}f \right)\p_y \nn
+&{\cal L}_{X_1} \eta_1 -{\cal L}_{Y_1} \xi_1 -\frac{1}{2}d \left( i_{X_1} \eta_1 - i_{Y_1} \xi_1\right) 
+\frac{1}{2}(df k -fdk-dg h+gdh) \nn
+&({\cal L}_{X_1}k -{\cal L}_{Y_1}h)dy.
\label{reduced Courant bracket}
\ena
By using the above formula (\ref{reduced Courant bracket}), 
the map $\mathcal{T}$ of the Courant bracket of the two basic vectors 
is given by exchanging $\p_y$ and $dy$ components as
\bea
&\mathcal{T}[X_1+ f\p_y +\xi_1 +h dy, Y_1+ g\p_y +\eta_1 +k dy]_C \nn
=&[X_1,Y_1] + ({\cal L}_{X_1}k -{\cal L}_{Y_1}h) \p_y \nn
+&{\cal L}_{X_1} \eta_1 -{\cal L}_{Y_1} \xi_1 -\frac{1}{2}d \left( i_{X_1} \eta_1 - i_{Y_1} \xi_1\right) 
+\frac{1}{2}(df k -fdk-dg h+gdh) \nn
+&\left({\cal L}_{X_1} g -{\cal L}_{Y_1}f \right) dy.
\ena
On the other hand, by exchanging $(f,h)$ and $(g,k)$ first 
in the same formula (\ref{reduced Courant bracket}), we have
\bea
&[\mathcal{T}(X_1+ f\p_y +\xi_1 +h dy), \mathcal{T}(Y_1+ g\p_y +\eta_1 +k dy)]_C \nn
=&[X_1+ h\p_y +\xi_1 +f dy, Y_1+ k\p_y +\eta_1 +g dy]_C \nn
=&[X_1,Y_1] + ({\cal L}_{X_1}k -{\cal L}_{Y_1}h) \p_y \nn
+&{\cal L}_{X_1} \eta_1 -{\cal L}_{Y_1} \xi_1 -\frac{1}{2}d \left( i_{X_1} \eta_1 - i_{Y_1} \xi_1\right) 
+\frac{1}{2}(dh g -hdg-dk f+kdf) \nn
+&\left({\cal L}_{X_1} g -{\cal L}_{Y_1}f \right) dy.
\ena
Thus, they are equivalent.
{\it (End of the proof.)}

\section{Reduction of the Koszul bracket \label{app:reduction}}
We here calculate that the Koszul bracket 
\bea
[\xi,\eta]_\theta
&=i_{\theta(\xi)}d\eta -i_{\theta(\eta)}d\xi +d(\theta (\xi,\eta )),
\ena
for general sections $\xi=\xi_1+fdy$ and $\eta=\eta_1+gdy$, that 
are not necessarily basic, and for a general Poisson tensor $\theta$.
We show that the Koszul bracket reduces to the form (\ref{reduced Koszul bracket}) if either of the following
conditions i) or ii) are satisfied:
\begin{enumerate}
\item[i)] $\theta$ is basic and $\theta_1=0$.
\item[ii)] $\theta$ is basic. The sections ($\xi$ and $\eta$) are basic and $\theta_1$-invariant.
\end{enumerate}
Our assumption in this paper is a particular case of i).

Note that
\bea
\theta (\xi)
&=\theta_2 (\xi_1) -f \theta_1+\theta_1 (\xi_1) \p_y ,\nn
\theta(\xi,\eta) 
&=\theta_2 (\xi_1,\eta_1 )-f\theta_1 (\eta_1) +g\theta_1 (\xi_1).
\ena
We divide the exterior differential of $M$ into $d=d_N +dy\p_y $ so that 
\bea
d\xi =d_N \xi_1+ dy \wedge {\cal L}_{\p_y} \xi_1 +d_N f\wedge dy.
\ena
Similarly, we have
\bea
d (\theta(\xi,\eta) )
&=d_N \left(\theta_2 (\xi_1,\eta_1 ) -f\theta_1 (\eta_1) +g\theta_1 (\xi_1) \right) 
+dy {\cal L}_{\p_y} \left(\theta_2 (\xi_1,\eta_1 ) -f\theta_1 (\eta_1) +g\theta_1 (\xi_1) \right),
\ena
and 
\bea
i_{\theta(\xi)}d\eta 
&=i_{\theta_2 (\xi_1) -f \theta_1+\theta_1 (\xi_1) \p_y}
(d_N \eta_1 +dy \wedge {\cal L}_{\p_y}\eta_1 + d_N g \wedge dy) \nn
&=i_{\theta_2 (\xi_1)}d_N \eta_1 
+ \left(  i_{\theta_2 (\xi_1) }d_N g -i_{\theta_2 (\xi_1) }{\cal L}_{\p_y}\eta_1\right) dy \nn
&\quad
-fi_{\theta_1}d_N \eta_1 
- f \left( i_{\theta_1}d_N g - i_{\theta_1}{\cal L}_{\p_y}\eta_1 \right) dy
+\theta_1 (\xi_1) \left( {\cal L}_{\p_y}\eta_1 - d_N g\right).
\ena
By using these, we compute
\bea
[\xi,\eta]_\theta
&=i_{\theta_2 (\xi_1)}d_N \eta_1 - i_{\theta_2 (\eta_1)}d_N \xi_1 
+ d_N \left(\theta_2 (\xi_1,\eta_1 )\right) \nn
&\quad -fi_{\theta_1}d_N \eta_1 +gi_{\theta_1}d_N \xi_1 
+d_N \left( -f\theta_1 (\eta_1) +g\theta_1 (\xi_1) \right) \nn
&\quad +\theta_1 (\xi_1) \left( {\cal L}_{\p_y}\eta_1 - d_N g\right)
-\theta_1 (\eta_1) \left( {\cal L}_{\p_y}\xi_1 - d_N f\right) \nn
&\quad + \left(  i_{\theta_2 (\xi_1) }d_N g  +\theta_1 (\xi_1){\cal L}_{\p_y} g
- i_{\theta_2 (\eta_1) }d_N f -\theta_1 (\eta_1){\cal L}_{\p_y} f \right)dy \nn
&\quad -\left( i_{\theta_2 (\xi_1) }{\cal L}_{\p_y}\eta_1 -i_{\theta_2 (\eta_1) }{\cal L}_{\p_y}\xi_1
+{\cal L}_{\p_y}( \theta_2 (\xi_1,\eta_1 ) )  \right) dy \nn
&\quad - f \left( i_{\theta_1}d_N g - i_{\theta_1}{\cal L}_{\p_y}\eta_1 
+{\cal L}_{\p_y}(\theta_1 (\eta_1))\right) dy 
+g \left( i_{\theta_1}d_N f - i_{\theta_1}{\cal L}_{\p_y}\xi_1 
+{\cal L}_{\p_y}(\theta_1 (\xi_1))\right) dy \nn
&=[\xi_1,\eta_1]^N_{\theta_2} \nn
&\quad -f{\cal L}^N_{\theta_1} \eta_1 +g{\cal L}^N_{\theta_1} \xi_1 
+\theta_1 (\xi_1) \left( {\cal L}_{\p_y}\eta_1 \right)
-\theta_1 (\eta_1) \left( {\cal L}_{\p_y}\xi_1 \right) \nn
&\quad + \left( {\cal L}_{\xi_1} g -{\cal L}_{\eta_1} f\right) dy 
 -({\cal L}_{\p_y}\theta_2 )(\xi_1,\eta_1 )  dy \nn
&\quad - f \left( {\cal L}^N_{\theta_1} g 
+({\cal L}_{\p_y}\theta_1 ) (\eta_1 )\right) dy 
+g \left( {\cal L}^N_{\theta_1} f 
+({\cal L}_{\p_y}\theta_1 )(\xi_1 ) \right) dy.
\label{decomposed Koszul}
\ena
Here in the second line, we defined $[\xi_1,\eta_1]^N_{\theta_2}=i_{\theta_2 (\xi_1)}d_N \eta_1 - i_{\theta_2 (\eta_1)}d_N \xi_1 + d_N \left(\theta_2 (\xi_1,\eta_1 )\right)$ and 
${\cal L}^N_{X_1}  =d_N i_{X_1}+ i_{X_1}d_N $.
We also used
\bea 
{\cal L}_{\xi_1} g
&=i_{\theta (\xi_1)}d g 
=i_{\theta_2 (\xi_1) +\theta_1 (\xi_1) \p_y}(d_N g+dy {\cal L}_{\p_y} g)\nn
&=i_{\theta_2(\xi_1)}d_N g +\theta_1 (\xi_1){\cal L}_{\p_y} g \nn
&={\cal L}^N_{\theta_2(\xi_1)} g +\theta_1 (\xi_1){\cal L}_{\p_y} g.
\label{Lxig}
\ena

From this result, we show the claim.
We first assume that only $\theta$ is basic. 
Then we may drop the terms involving ${\cal L}_{\p_y}$ acting on $\theta_2$ and $\theta_1$, 
and we obtain 
\bea
[\xi,\eta]_\theta
&=[\xi_1,\eta_1]^N_{\theta_2} \nn
&\quad -f{\cal L}^N_{\theta_1} \eta_1 +g{\cal L}^N_{\theta_1} \xi_1 
+\theta_1 (\xi_1) \left( {\cal L}_{\p_y}\eta_1 \right)
-\theta_1 (\eta_1) \left( {\cal L}_{\p_y}\xi_1 \right) \nn
&\quad + \left( {\cal L}_{\xi_1} g -{\cal L}_{\eta_1} f\right) dy  
 - f \left( {\cal L}^N_{\theta_1} g 
\right) dy 
+g \left( {\cal L}^N_{\theta_1} f 
\right) dy.
\label{decomposed Koszul with basic theta}
\ena
It reduces to the desired form 
\bea
[\xi,\eta]_\theta
&=[\xi_1,\eta_1]^N_{\theta_2} 
+ \left( {\cal L}_{\xi_1} g  -{\cal L}_{\eta_1} f\right) dy,
\ena
if $\theta_1=0$ (note that ${\cal L}_{\xi_1} g={\cal L}^N_{\theta_2(\xi_1)} g$). 
This is the condition i).

Next , let us assume that $\xi$, $\eta$ and $\theta$ are all basic.
Then we have
\bea
[\xi,\eta]_\theta
&=[\xi_1,\eta_1]^N_{\theta_2} 
 -f{\cal L}^N_{\theta_1} \eta_1 +g{\cal L}^N_{\theta_1} \xi_1 \nn
&\quad + \left( {\cal L}^N_{\theta_2(\xi_1)} g  -{\cal L}^N_{\theta_2(\eta_1)} f \right) dy 
 - f \left( {\cal L}^N_{\theta_1} g 
\right) dy 
+g \left( {\cal L}^N_{\theta_1} f 
 \right) dy.
\ena
it reduces to the desired form iff
\bea
{\cal L}^N_{\theta_1} \xi_1=0, ~~{\cal L}^N_{\theta_1}\eta_1=0,~~
{\cal L}^N_{\theta_1} f=0, ~~{\cal L}^N_{\theta_1}g =0.
\ena
These are satisfied either $\theta_1=0$ (a particular case of i)), 
or ii) sections are $\theta_1$-invariant.

Next, we consider the case of the Dirac structure $L={\rm span}\{dx^m, \p_y \}\ni \xi_1 + f\p_y$.
The bracket of the Courant algebroid $(TM)_0\oplus (T^*M)_\theta$ for 
sections of $L$ is given by
\bea
[\xi_1+f\p_y, \eta_1+g\p_y]
=[\xi_1,\eta_1]_\theta + {\cal L}_{\xi_1}(g\p_y) -{\cal L}_{\eta_1}(f\p_y)
+\textstyle{\frac{1}{2}}d_\theta (i_{f\p_y}\eta_1 -i_{g\p_y}\xi_1 ).
\ena 
We assume that only $\theta$ is basic.
Then, the first term is written by using (\ref{decomposed Koszul with basic theta}) as
\bea
[\xi_1,\eta_1]_\theta
&=[\xi_1,\eta_1]^N_{\theta_2}
+\theta_1 (\xi_1) \left( {\cal L}_{\p_y}\eta_1 \right)
-\theta_1 (\eta_1) \left( {\cal L}_{\p_y}\xi_1 \right)
\ena
The second term is written as (see (\ref{Lxig}))
\bea
{\cal L}_{\xi_1}(g\p_y)
&=({\cal L}_{\xi_1}g)\p_y -g i_{\xi_1} {\cal L}_{\p_y}\theta  \nn
&=({\cal L}_{\xi_1}g)\p_y.
\ena
Thus, we obtain
\bea
[\xi_1+f\p_y, \eta_1+g\p_y]
=[\xi_1,\eta_1]^N_{\theta_2}
+\theta_1 (\xi_1) \left( {\cal L}_{\p_y}\eta_1 \right)
-\theta_1 (\eta_1) \left( {\cal L}_{\p_y}\xi_1 \right)
+({\cal L}_{\xi_1}g - {\cal L}_{\eta_1}f)\p_y.
\ena 
It is again an element of $\Gamma(L)$ either if $\theta_1=0$ (condition i)), 
or if $\xi_1, \eta_1, f$ and $g$ are all basic (the condition ii)).
In both cases, the bracket reduces to the desired form.
Therefore, the condition i) or ii) is sufficient to prove the T-duality.



\end{document}